\documentclass{aastex631}

\usepackage{color}

\graphicspath{{./}{figures/}}

\begin{document}

\title{Two high-amplitude $\delta$ Scuti-$\gamma$ Doradus hybrids constrained by the radial fundamental p and equally spaced g modes}
\correspondingauthor{Xinghao Chen}

\email{chenxinghao1003@163.com}
%
\author[0000-0003-3112-1967]{Xinghao Chen}
\affiliation{Yunnan Observatories, Chinese Academy of Sciences, P.O. Box 110, Kunming 650216, China}
\affiliation{Key Laboratory for Structure and Evolution of Celestial Objects, Chinese Academy of Sciences, P.O. Box 110, Kunming 650216, China}
\affiliation{International Centre of Supernovae, Yunnan Key Laboratory, Kunming 650216, P. R. China}
\author[0000-0002-5164-3773]{Xiaobin Zhang}
\affiliation{Key Laboratory of Optical Astronomy, National Astronomical Observatories, Chinese Academy of Sciences, Beijing, 100012, China}
\author{Yan Li}
\affiliation{Yunnan Observatories, Chinese Academy of Sciences, P.O. Box 110, Kunming 650216, China}
\affiliation{Key Laboratory for Structure and Evolution of Celestial Objects, Chinese Academy of Sciences, P.O. Box 110, Kunming 650216, China }
\affiliation{University of Chinese Academy of Sciences, Beijing 100049, China}
\affiliation{Center for Astronomical Mega-Science, Chinese Academy of Sciences, 20A Datun Road, Chaoyang District, Beijing, 100012, China}
\author{Jie Su}
\affiliation{Yunnan Observatories, Chinese Academy of Sciences, P.O. Box 110, Kunming 650216, China}
\affiliation{Key Laboratory for Structure and Evolution of Celestial Objects, Chinese Academy of Sciences, P.O. Box 110, Kunming 650216, China}
\affiliation{International Centre of Supernovae, Yunnan Key Laboratory, Kunming 650216, P. R. China}

\begin{abstract}
Based on 2-minute cadence TESS data, we investigate pulsations of TIC 65138566 and TIC 139729335 and discover them to be two new HADS stars with equally spaced g modes. We recognize the radial fundamental mode $f_1$ = 18.3334 c/d and the first overtone $f_3$ = 23.6429 c/d for TIC 65138566, and identify the highest peak $f_1$ = 19.0955 c/d as the radial fundamental mode for TIC 139729335. For g modes, both stars display a regular period spacing of 2413 s. Through detailed seismological analysis, we deduce that these period spacing patterns correspond to modes with $\ell$ = 1. Moreover, our analysis reveals that with the increase in masses and metallicities, the star should display a higher degree of evolution to match a specific period spacing $\Pi_0$. Conversely, the star should have a lower extent of evolution to match the radial fundamental mode. These two contradictory behaviors allow us to precisely obtain stellar physical parameters.  TIC 65138566 and TIC 139729335 are determined to be two main sequence stars that have almost the same range of masses and metallicities, with $M$ = 1.36 $\pm$ 0.06 $M_{\odot}$ and $Z$ = 0.005 $\pm$ 0.002. The hydrogen abundance in the core of TIC 65138566 is estimated to be about 0.28, while TIC 139729335 has a slightly higher value of around 0.31. Finally, we suggest that the HADS $\delta$ Scuti-$\gamma$ Doradus star TIC 308396022 is a main sequence star with $M$ = 1.54 $\pm$ 0.08 $M_{\odot}$, $Z$ = 0.007 $\pm$ 0.001, and $X_{\rm c}$ = 0.18 $\pm$ 0.02. 

\end{abstract}


\section{Introduction}
The $\delta$ Scuti and $\gamma$ Doradus variables are two groups of A- to F-type stars that located on or near the main sequence evolutionary stage. The $\delta$ Scuti stars mainly pulsate in high-frequency oscillations with periods ranging from 0.02 to 0.25 days (Breger 2000). In terms of their amplitudes, these stars can be classified into two subgroups: the high-amplitude $\delta$ Scuti stars (HADS), whose amplitudes approximate or surpass 0.1 magnitude (Petersen 1989; Petersen \& Christensen-Dalsgaard 1996), and the low-amplitude $\delta$ Scuti stars (LADS). For LADS, a large number of pulsation frequencies have been observed. However, it is very difficult to identify them since the asymptotic theory is invalid due to the low radial order. Conversely, identification of the modes in HADS is clear and straightforward because HADS typically pulsate in the radial fundamental mode and/or low overtone radial modes (McNamara 2000), such as for VX Hydrae (Xue et al. 2018), GSC 4552-1498 (Sun et al. 2022), KIC 2987323 (Yang et al. 2022), and TIC 448892817 (Lv et al. 2022). The pulsation of $\delta$ Scuti stars contains both radial and nonradial p modes, signifying the potential to investigate the stellar envelope. The $\gamma$ Doradus stars mainly pulsate in low-degree and high-order g modes with periods in the range of 0.3 and 3 days. The g modes in $\gamma$ Doradus stars provide us with an opportunity to insight into the deeper interiors of the star. In addition, it is noteworthy that the asymptotic theory is valid for these high-order g modes and predicts a uniform period spacing (Unno et al. 1979; Tassoul 1980). The approximately constant period spacings and the deviations from these spacings have been used to investigate the physics of the stellar interior, such as chemical mixing and rotation (Miglio et al. 2008; Van Reeth et al. 2015, 2016, 2018).

The instability strip of $\delta$ Scuti and that of the $\gamma$ Doradus stars overlap in part in the Hertzsprung–Russell diagram (Balona 2011; Henry et al. 2011; Uytterhoeven et al. 2011; Xiong et al. 2016). Accordingly, stars situated in the overlapping region are expected to manifest both the high-frequency p mode features of $\delta$ Scuti stars and the low-frequency g mode oscillations of $\gamma$ Doradus stars (Xiong et al. 2016). Henry \& Fekel (2005) discovered the first such $\delta$ Scuti-$\gamma$ Doradus hybrid star from observation on the ground. Subsequently, HD 49434 and HD 8801 were also identified (Uytterhoeven et al. 2008; Handler 2009). Due to the space missions CoRoT (Baglin et al. 2006) and Kepler (Broucki et al. 2010), the hybrid behaviors have been discovered to be prevalent among stars with spectral types between A and F (Grigahc\'ene et al. 2010; Hareter et al. 2010; Balona et al. 2015). Numerous $\delta$ Scuti-$\gamma$ Doradus hybrid pulsators have been observed and identified (Uytterhoeven et al. 2011, Bradley et al. 2015). The hybrid stars are greatly significant since the coexistence of p and g modes offers us an opportunity to investigate the structure and evolution of the star, from the envelope to the inner core. Seismological modeling has been conducted on various hybrid $\delta$ Scuti-$\gamma$ Doradus stars, such as for CoRoT 105733033 (Chapellier et al. 2012), KIC 11145123 (Kurtz et al. 2014), KIC 9244992 (Saio et al. 2015), HD 49434 (Brunsden et al. 2015), and CoRoT 100866999 ( Chapellier et al. 2013). Thereafter, more detailed seismological modelings for these five stars were performed by S\'anchez Arias et al. (2017), and Chen et al. (2019) estimated the size of the convective core of CoRoT 100866999 to be about 9.3\% of the stellar radius. 

Recently, the first HADS with regular period-spaced g modes was discovered by Yang et al. (2021) from the TESS photometric data. Such star is very interesting since their $\delta$ Scuti-type oscillations may be easily recognized. TIC 65138566 (TYC 6532-2202-1; $\alpha_{2000}$ = $07^h10^m58^s.350$,  $\delta_{2000}$ =  $-27^\circ21'00''.463$; V = 11.19 mag) and TIC 139729335 (TYC 8435-371-1; $\alpha_{2000}$ = $21^h32^m33^s.885$,  $\delta_{2000}$ =  $-52^\circ20'49''.248$; V = 11.54 mag) are found to be two HADS stars similar to TIC 308396022. Therein, TIC 65138566 was identified to be a HADS star in the catalog of Barac et al. (2022), whereas TIC 139729335 is a newly identified object. In this work, we aim to conduct a comprehensive analysis for the two stars TIC 65138566 and TIC 139729335. Section 2 presents the observation data, and Section 3 focus on the extractions and identifications of their respective frequencies. Subsequently, in Section 4, we elaborate on our asteroseismic modeling. Finally, we summarize and discuss our results in Section 5.
   
\section{Observations and data reduction}
TIC 65138566 was observed by the TESS satellite in 2-minute cadence mode during Sectors 33 and 34 from 2021 January 13 to 2021 February 9. The TESS photometric observations for TIC 139729335 include Sector 1 from 2018 July 25 to 2018 August 22 in the 30-minute cadence mode, and Sectors 27 and 28 from 2020 July 30 to 2020 August 26 in the 2-minute cadence mode. In this study, only the 2-minute cadence data were used. We downloaded the photometric data from the Mikulski Archive for Space Telescopes(MAST,https://mast.stsci.edu/portal/Mashup/Clients/Mast/Portal.html). The data we used for analysis include the times in Barycentric Julian Date (BJD) format and the "PDCSAP$\_$FLUX" fluxes. The "PDCSAP$\_$FLUX" values have been processed by the Pre-search Data Conditioning Pipeline (Jenkins et al.2016) in order to remove instrumental trends. To prepare the data for analysis, we removed the outliers from the light curves. Then, we normalized the fluxes using the method described by Slawson et al. (2011). Figure 1 displays the resulting light curves for TIC 65138566 and TIC 139729335. Both stars display amplitudes of around 0.15 mag in the figure.

\section{Frequency analysis}
To extract pulsation frequencies, we conducted a multiple-frequency analysis of the photometric data with the Period04 program (Lenz \& Breger, 2005). Following Breger et al. (1993), the empirical threshold of S/N = 4 was adopted for a frequency that is accepted to be significant. The significant peaks are extracted one by one through a pre-whitening method. A total of 72 frequencies with S/N $\geq$ 4 were extracted for TIC 65138566, while 87 frequencies with S/N $\geq$ 4 were obtained for TIC 139729335. The amplitude spectra are presented in Figure 2. The extracted frequencies of TIC 65138566 can be found in Table 1, and those of TIC 139729335 are listed in Table 2. In both tables, noises were computed within a range of 2 day$^{-1}$ around each frequency, and the uncertainties in frequencies and amplitudes were determined using Monte Carlo simulations as proposed by Fu et al. (2013).

Following P{\'a}pics et al. (2012) and Kurtz et al. (2015), we carefully examine the extracted frequencies in the form of $f_i = mf_j + nf_k$ to eliminate possible linear combinations. A peak is recognized as a combination frequency if the amplitudes of the combination terms are significantly lower compared to their parent frequencies, meanwhile, the discrepancy between the predicted frequency $f_i$ and the observed frequency is smaller than the frequency resolution 1.5/$\Delta$T (Loumos $\&$ Deeming 1978). Additionally, high-order combinations are considered only when the lower-order terms are detected (P{\'a}pics et al. 2012). A total of 54 combination frequencies are identified for TIC 65138566, whereas 67 combination frequencies are distinguished for TIC 139729335. Finally, 18 confident independent frequencies, including 7 p modes and 11 g modes, are retained for TIC 65138566. As for TIC 139729335, 20 confident independent frequencies remain, including 11 p modes and 9 g modes. 

\subsection{The $\delta$ Scuti domain}
In the $\delta$ Scuti-type frequency region, there are 7 confident frequencies in the range [18.3334; 30.3728] day$^{-1}$ for TIC 65138566 and 11 confident frequencies in the interval [15.2687; 49.8617] day$^{-1}$ for TIC 139729335. Figure 2 depicts that the amplitude spectra of TIC 65138566 and TIC 139729335 are largely dominated by their respective highest peaks. In the case of high-amplitude $\delta$ Scuti stars, it is commonly presumed that the highest peak corresponds to the radial fundamental mode. For TIC 65138566, the dominant frequency $f_1$ = 18.3334 day$^{-1}$ has an amplitude of 35.506 mmag. Furthermore, the period ratio between $f_1$ and $f_3$ is 0.775, which agrees perfectly with the result of Petersen \& Christensen-Dalsgaard (1996). Hence, we suggest that $f_1$ and $f_3$ are the radial fundamental mode and the radial first overtone mode, respectively. For TIC 139729335, the highest amplitude frequency $f_1$ = 19.0955 day$^{-1}$ has an amplitude of 46.625 mmag. The period ratio between $f_1$ and $f_{14}$ is 0.764, and the period ratio between $f_2$ and $f_{14}$ is 0.778. Both ratios are consistent with the result of Petersen \& Christensen-Dalsgaard (1996). Given that the amplitude of $f_1$ is approximately four times that of $f_2$, we propose that the frequency $f_1$ is the radial fundamental mode.

Moreover, we find remarkable coupling features between the $\delta$ Scuti frequencies and the $\gamma$ Doradus frequencies for TIC 65138566 and TIC 139729335, such as $f_{23} = p_1-g_1$, $f_{25}= F-g_3$, $f_{26}= p_1-g_2$, and $f_{27}= F-g_1$ for TIC 65138566, and $f_{15}= F-g_1$, $f_{16}= F-g_2$, $f_{17}= p_1+g_6$, and $f_{18}= F+g_1$ for TIC 139729335. Kennedy et al.(1993) predicted such couplings between p modes and g modes for the Sun. Chapellier \& Mathias (2013) pointed out that the couplings may be a common character of hybrid $\delta$ Scuti-$\gamma$ Doradus stars.  Detections of these couplings further confirm that the $\delta$ Scuti frequencies and the $\gamma$ Doradus frequencies that we detected arise from the same star.

\subsection{The $\gamma$ Doradus domain}
G-modes, known as gravity waves, can propagate in the star deeper than p modes. We extracted 11 confident g modes in the range [1.1192; 2.1112] day$^{-1}$ for TIC 65138566, and 9 confident g modes in the interval [0.9656; 4.0286] day$^{-1}$ for TIC 139729335. For high-order g modes, the asymptotic theory predicts a nearly uniform period spacing between modes with the same degree $\ell$ but consecutive radial orders $n$ (Unno et al. 1979; Tassoul 1980), i.e.,
\begin{equation}
\Delta P(\ell) = P_{\ell,n+1}-P_{\ell,n} \approx \frac{\Pi_{0}}{\sqrt{\ell(\ell+1)}} = \frac{2\pi^{2}(\int_{0}^{R}\frac{N}{r}dr)^{-1}}{\sqrt{\ell(\ell+1)}}.
\end{equation}
In the equation, $P_{\ell,n}$ is the period of the oscillation mode characterized by the indices $\ell$ and $n$, $N$ is the Brunt-V$\ddot{a}$is$\ddot{a}$l$\ddot{a}$ frequency, and $\Pi_0$ represents the period spacing that is independent of $\ell$. To search for equivalent period spacing patterns, we conducted a Kolmogorov-Smirnov (KS) test (Kawaler 1988) on these independent $\gamma$ Doradus frequencies. The results are depicted in the left panels of Figure 3. The left panels in Figure 3 show the probability of a given set of period spacings $\Delta$P, covering a range from 100 to 4000 seconds. In these panels, the lowest logarithmic value of Q signifies the statistical significance of the period spacing. For TIC 65138566, there is a prominent minimum value of $\log Q$ at period spacing $\Delta P$ = 2417.6 s. For TIC 139729335, the minimum $\log Q$ value is found at $\Delta P$ = 2414.1 s. Afterwards, we examine potential period-spacing patterns, identifying 10 out of the 11 g modes of TIC 65138566 in a period-spacing pattern, and 7 out of 9 g modes of TIC 139729335 in a period-spacing pattern. The residual modes may be oscillations of other spherical harmonic degree $\ell$. It is difficult to identify them due to too few in number. Table 3 presents details of these pulsations in order of increasing period. In the right panels of Figure 3, we display the period $P_n$ as a function of the radial order $n$,  where $n$ is the integer around $P_n/ \Delta P$.  The least-squares analysis yields a rigorous linear relation of $P_n$ = 2412.7($\pm$5.7) $\times$ n + 25.9($\pm$137.0) s for TIC 65138566 and $P_n$ = 2412.7($\pm$4.0) $\times$ n + 261.6($\pm$110.9) s for TIC 139729335, respectively. The nearly constant period spacings may indicate a very low near-core rotation rate for TIC 65138566 and TIC 139729335. Splittings and period spacing patterns of numerous $\gamma$ Doradus pulsators have been investigated (Kurtz et al. 2014; Ouazzani et al. 2017; Saio et al. 2018; Li et al. 2019a, 2019b, 2020). Kurtz et al. (2014) detected rotationally split core g-mode triplets and surface p-mode triplets and quintuplets in the star KIC 11145123, providing the first robust determination of the rotation of the deep core and surface of a main-sequence star. Li et al. (2019a) analyzed 22 slowly rotating $\gamma$ Doradus stars and observed rotational splittings in almost all of these stars. However, for TIC 65138566 and TIC 139729335, the identified g modes seem to be singular, as there is no detection of rotationally split multiplets. This may be attributed to the relatively short time span of the TESS photometric data, which hinders the differentiation of rotational splittings. Additionally, nearly all of the g modes can be found in period spaced patterns. This slow rotation is also consistent with the observation that HADS stars usually rotate slowly (McNamara 2000).

\section{Asteroseismic models}
The stellar models used in this study were computed with the one-dimensional stellar evolution code called Modules for Experiments in Stellar Astrophysics (MESA, version 10398, Paxton et al. 2011, 2013, 2015, 2018). The corresponding adiabatic frequencies were calculated using a MESA submodule called “pulse-adipls” (Christensen-Dalsgaard 2008). In the calculations,  the OPAL equation of state tables (Rogers \& Nayfonov 2002) were adopted.  The OPAL opacities from Iglesias \& Rogers 1996 were used in the high temperature region,  while those from Ferguson et al. (2005) were adopted in the low temperature region.  The initial metal ingredient used in this work was the solar metal composition AGSS09 (Asplund et al.  2009).  Additionally,  we applied "simple$\_$photosphere" as the atmospheric boundary conditions and the classical mixing length theory (B$\ddot{\rm o}$hm-Vitense 1958) with the solar value of $\alpha$ = 1.9 (Paxton et al. 2013) to treat the convective.  Besides,  our models did not account for the effects of element diffusion, rotation, and magnetic field on stellar structure and evolution.

In this work, each star evolves from the zero-age main sequence to the evolutionary stage where the hydrogen in the core is depleted. The initial stellar masses considered range from 1.3 to 2.5 $M_{\odot}$ with a step of $\Delta M$ = 0.02 $M_{\odot}$. The initial metallicity $Z$ is varied between 0.001 and 0.030 with a step of $\Delta Z$ = 0.001. The initial helium abundance $Y$ considered in the present work is related to $Z$ through the relation $Y$ = 0.249 + 1.5 $Z$ (Choi et al. 2016). For the overshooting mixing of the convective core, a diffusion coefficient that exponentially decays (Freytag et al. 1996; Herwig 2000) is adopted. Guo \& Li proposed that the overshooting parameters $f_{\rm ov}$ should be approximately 0.01 for low-mass main sequence stars. Therefore, we set the overshooting parameter $f_{\rm ov}$ to a fixed value of 0.01 in this study.

Figure 4 illustrates evolutions of the period spacing $\Pi_0$ and the radial fundamental mode $F$ as a function of normalized central hydrogen abundance $X_{\rm c}/X_{0}$, where $X_0$ denotes the initial hydrogen abundance of the star. In the figures, $X_{\rm c}/X_{0}$ = 1 corresponds to the zero-age main sequence stage and $X_{\rm c}/X_{0}$ = 0 corresponds to the terminal-age main sequence stage. The upper pane resembles Figure 1 from Mombarg et al. (2019), but with different parameters. It can be seen that values of the asymptotic period spacing $\Pi_0$ decrease as the star evolves. Besides, the figure implies that the frequencies in the identified period-spacing patterns correspond to modes with $\ell$ = 1. For the radial fundamental mode $F$, its value decreases as the star evolves until the central hydrogen abundance $X_{\rm c}$ reaches approximately 0.05. Subsequently, the value slightly increases until the central hydrogen is exhausted, before decreasing again. Additionally, the upper panel illustrates that as the masses and metallicities increase, the star should display a higher degree of evolution (i.e., less central hydrogen in the core) to reproduce a given period spacing $\Pi_0$. On the other hand, contrasting with $\Pi_0$, the lower panel displays that as the masses and metallicities  increase, the star should show a lower degree of evolution (i.e., more central hydrogen in the core) to match a specific frequency of the radial fundamental mode. The two conflicting inclinations may allow us to accurately determine their physical parameters. Mombarg et al. (2019) stated that obtaining stellar parameters with relative uncertainties of approximately 10 percent is a cumbersome and challenging task, even with high-precision spectroscopic measurements and $\Pi_0$. Moreover, they posited that including more types of detected modes can potentially enhance the accuracy of these parameters. Based on the above analysis, it may be appropriate to consider the inclusion of the radial fundamental mode. For TIC 65138566 and TIC 139729335, we obtain values of the period spacing and the radial fundamental mode simultaneously. In Figure 4, we mark $\Pi_0$ and $F$ of TIC 65138566 and TIC 139729335 with horizontal dashed lines. Moreover, we deduced that TIC 65138566 and TIC 139729335 are two main sequence stars from the figures.

According to the TESS input catalog, TIC 65138566 and TIC 139729335 have an effective temperature of 7487 K and 7638 K, respectively. In this study, we adopt a slightly wider range of $T_{\rm eff}$ = 7500 $\pm$ 500 K as the observed constraint. We calculate the adiabatic frequencies of the radial fundamental mode( $\ell$ = 0 and $n$ = 0) for all evolutionary models that fall inside the above constraint. 

Figure 5 depicts changes of $\Pi_0$ versus various physical parameters, where panels (a)-(c) are for TIC 65138566 and panels (d)-(f) are for TIC 139729335. Therein, each square represents one optimal fitting model for the radial fundamental mode throughout its evolutionary track. From the figure, it is apparent that determining the parameters and evolutionary stages of TIC 65138566 and TIC 139729335 solely based on the radial fundamental mode is very difficult. To obtain precise parameters and evolutionary stages, additional information is required. The horizontal lines demarcate the period spacing ranges of 3412 $\pm$ 100 s. The red squares denote the preferred models that fall inside this period spacing range. When the period spacing $\Pi_0$ is taken into account, the parameters and evolutionary stages can be accurately determined. Table 4 provides the details of these preferred models, and Table 5 presents the physical parameters of TIC 65138566 and TIC 139729335 determined by $F$ and $\Pi_0$. TIC 65138566 and TIC 139729335 are two main sequence stars with almost the same range of masses and metallicties ($M$ = 1.36 $\pm$ 0.06 $M_{\odot}$ and $Z$ = 0.005 $\pm$ 0.002). It is worth mentioning that TIC 65138566 is more evolved than TIC 139729335, and it has a larger radius but a comparatively smaller convective core. The present central hydrogen abundance of TIC 65138566 is about 0.28, while TIC 139729335 has a slightly higher value of around 0.31.
\section{Summary and Discussion}
Based on 2-minute cadence TESS photometric data, we investigated the pulsations of TIC 65138566 and TIC 139729335 and identified them as two new HADS stars companied by regular period-spaced g modes. We recognized the radial fundamental mode $f_1$ and the radial first overtone $f_3$ for TIC 65138566. For TIC 139729335, the most prominent peak $f_1$ is suggested to be the radial fundamental mode. In the low-frequency region, the KS tests revealed that the g modes of both stars exhibit equally spaced period patterns with a period spacing of $\Delta P$ = 2412.7 s. 

We analyzed the evolutionary behaviors of the asymptotic spacing $\Pi_0$ and the radial fundamental mode $F$. Our analysis reveals that as the masses and metallicities increase, the star is expected to exhibit a higher degree of evolution to match a specific period spacing $\Pi_0$. In contrast, the star should display a lower degree of evolution to match a given frequency of the radial fundamental mode. The two conflicting tendencies enable us to accurately determine the physical parameters of the star.For TIC 65138566 and TIC 139729335, we computed a grid of models to match their asymptotic spacing $\Pi_0$ and the radial fundamental mode $F$. Despite using a moderately broad range of observed constraints, specifically $T_{\rm eff}$ = 7500 $\pm$ 500 K, the parameters of both stars can still be accurately determined by $F$ and $\Pi_0$. The fitting results show that TIC 65138566 and TIC 139729335 are two main sequence stars with almost the same range of masses and metallicities, i.e., $M$ = 1.36 $\pm$ 0.06 $M_{\odot}$ and $Z$ = 0.005 $\pm$ 0.002. Besides, we find that TIC 65138566 is more evolved than TIC 139729335. The present hydrogen abundance in the core of TIC 65138566 is about 0.28, while TIC 139729335 has a slightly higher value of around 0.31. The mass of $\delta$ Scuti stars generally ranges from 1.5 to 2.5 $M_{\odot}$ (Breger et al. 2000; Aerts et al. 2010), while the masses of $\gamma$ Doradus stars are in the range of 1.4-2.0 $M_{\odot}$ (Dupret et al. 2004; Xiong et al. 2016; Van Reeth et al. 2016). However, the masses of both TIC 65138566 and TIC 139729335 are lower than the typical values.  We carefully examine their models and find that their structures are quite normal, i.e., a contracting convective core and an expanding envelope. This observed low mass may only be a consequence of poor metallicity.

Figure 6 illustrates plots of $\Pi_0$ versus the average density $\sqrt{\bar{\rho}/\bar{\rho}_{\odot}}$ and the relative radius of the convective core $R_{\rm conv}/R$. Panels (a) and (b) are for TIC 65138566, while panels (c) and (d) are for TIC 139729335. From panels (a) and (c), it can be seen that the average densities of both stars are well determined by the radial fundamental mode. It is easy to understand this according to the basic pulsation relation
\begin{equation}
P\sqrt{\bar{\rho}/\bar{\rho}_{\odot}}=Q,
\end{equation}
where $P$ denotes the pulsation period, $\bar{\rho}$ represents the average density, and $Q$ is the pulsation constant. For the particular radial fundamental mode, the $Q$ value remains constant for $\delta$ Scuti stars, i.e., $Q$ =0.033 day. In light of our previous discussion, the parameters and evolutionary stages of TIC 65138566 and TIC 139729335 can not be determined precisely solely based on the radial fundamental mode, unless the period spacing $\Pi_0$ is also taken into account. The radial fundamental mode mainly propagate in the stellar envelope, while these g modes can propagate in the deep interior of the star.
In panels (b) and (d), it can be seen that $\Pi_0$ show a notable liner correlation with $R_{\rm conv}/R$. When we consider $\Pi_0$, the size of the convective core for TIC 65138566 and TIC 139729335 are estimated to be around 6.8\% and 7.1\% of the stellar radius, respectively. 

In this study, we have implemented an adiabatic analysis based on the MESA submodule called "pulse-adipls". Our approach uses the radial fundamental mode and the period spacing as constraints to determine the parameters and evolution stages of TIC 65138566 and TIC 139729335. It is important to note that we have not compared g modes with their theoretical counterparts one by one. Based on Model 6 of TIC 65138566 in Table 4, we find the non-adiabatic code GYRE (Townsend \& Teitler 2013; Townsend et al. 2018; Goldstein \& Townsend 2020) does not influence the period spacing, while predicting a slightly higher value of 18.503 c/d for the radial fundamental mode compared to the adiabatic value of 18.335 c/d. This subtle discrepancy only has a minor impact on our conclusions, resulting in a slightly reduced extent of evolution. Additionally, the non-adiabatic code indicates instability in the radial modes of both TIC 65138566 and TIC 139729335 while suggesting stability in the observed g modes among the hybrids. The excitation mechanism of g modes in $\delta$ Scuti variables remains incompletely understood, thus necessitating further investigation in future studies.

Finally, TIC 308396022 is another high-amplitude $\delta$ Scuti star with regular g-mode period spacing discovered by Yang et al. (2021). Yang et al. (2021) recognized the strongest peak 13.20363 c/d as the radial fundamental mode and established the asymptotic period spacing as $\Pi_0$ = 3655 s. Based on the evolutionary models in this work, TIC 308396022 is inferred to be a main sequence star with $M$ = 1.54 $\pm$ 0.08 $M_{\odot}$, $Z$ = 0.007 $\pm$ 0.001, $T_{\rm eff}=7367^{+438}_{-348}$ K, $\log (L/L_{\odot})$ = 1.04 $\pm$ 0.12, $\log g$ = 4.00 $\pm$ 0.01, $R$ = 2.04 $\pm$ 0.04 $R_{\odot}$, $X_{\rm c}$ = 0.18 $\pm$ 0.02, $\bar{\rho}$ = 0.180 $\pm$ 0.002 $\bar{\rho}_{\odot}$, and $R_{\rm conv}/R$ = 0.059 $\pm$ 0.002.    

\begin{acknowledgments}
We are sincerely grateful to the anonymous referee for instructive advice and productive suggestions. This work is supported by the National Key R\&D Program of China (Grant No.2021YFA1600400/2021YFA1600402) and by the B-type Strategic Priority Program No. XDB41000000 funded by the Chinese Academy of Sciences. The authors also appreciate the support of the National Natural Science Foundation of China (grant NO. 12173080 to X-H.C., 12373037 to X-B.Z., 12133011 and 12288102 to Y.L.). X-H.C. also sincerely acknowledges the supports of the Yunnan Revitalization Talent Support Program Young Talent Project, the Yunnan Fundamental Research Projects (202101AT070006), and the Youth Innovation Promotion Association of the Chinese Academy of Sciences,  as well as the International Centre of Supernovae Yunnan Key Laboratory (No. 202302AN360001).X-B.Z. also acknowledges the support of the science research grants from the China Manned Space Project with NO. CMS-CSST-2021-A10. J.Su acknowledges the support of Light of West China Program.   This paper includes data collected with the TESS mission, obtained from the MAST data archive at the Space Telescope Science Institute (STScI). Funding for the TESS mission is provided by the NASA Explorer Program. STScI is operated by the Association of Universities for Research in Astronomy, Inc., under NASA contract NAS 5–26555. The authors sincerely acknowledge them for providing such excellent data.

\end{acknowledgments}

\begin{figure*}
\centering
\includegraphics[width=0.5\textwidth, angle = 0]{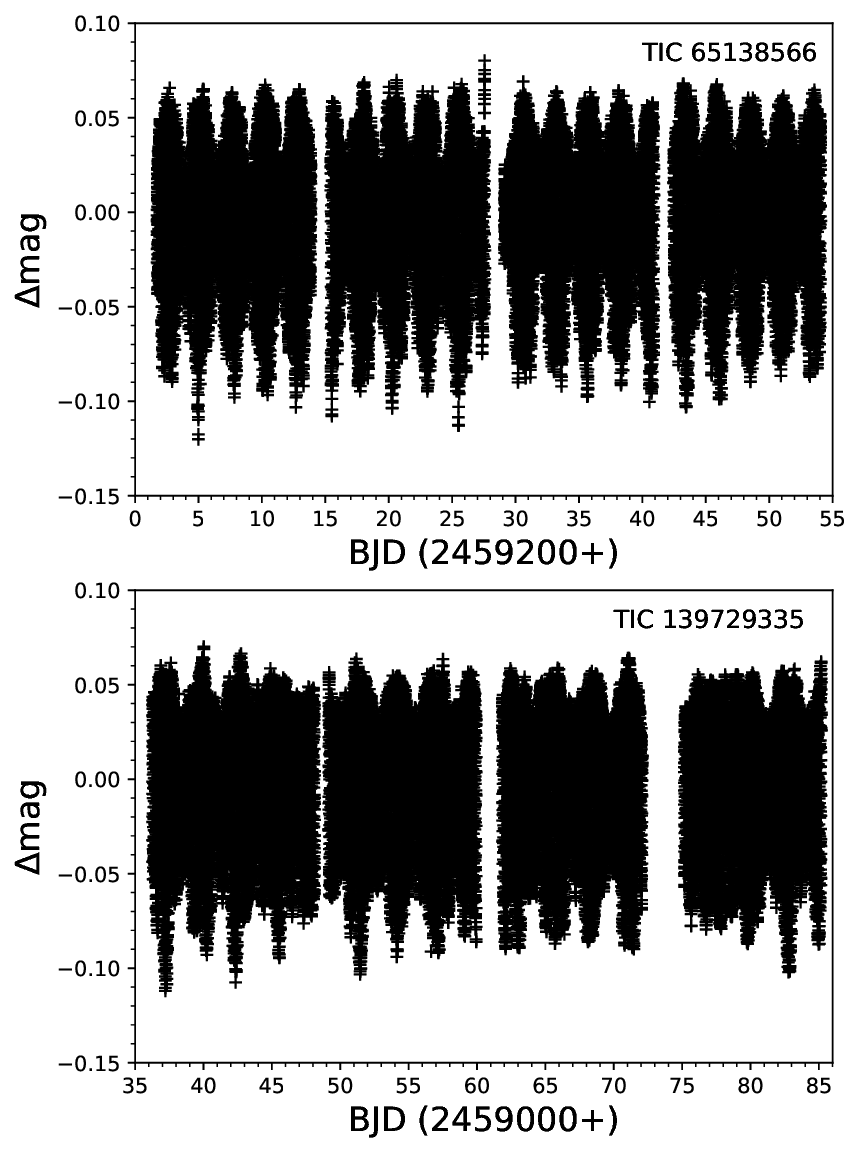}
  \caption{\label{Figure 1} Light curves.}
\end{figure*}

\begin{figure*}
\gridline{\fig{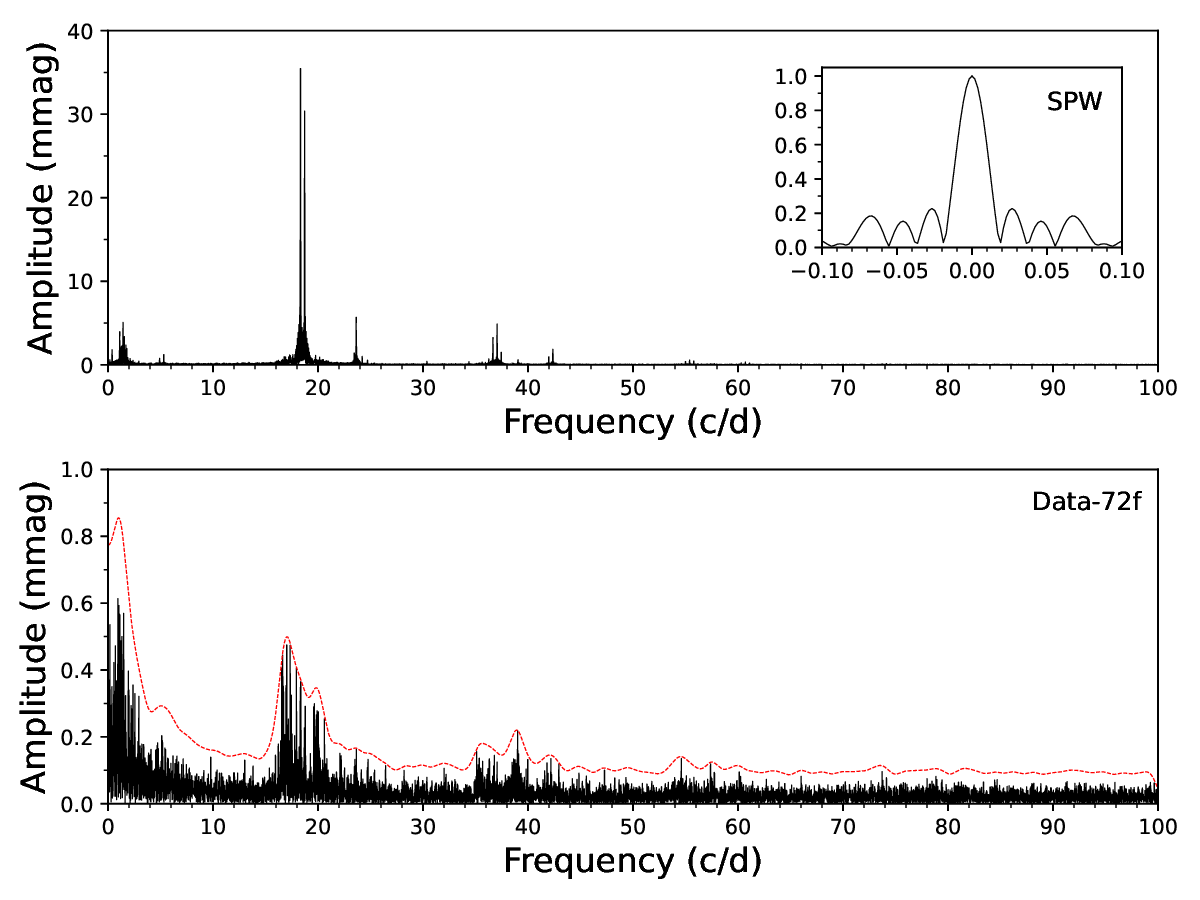}{0.5\textwidth}{(a) TIC 65138566}
 \fig{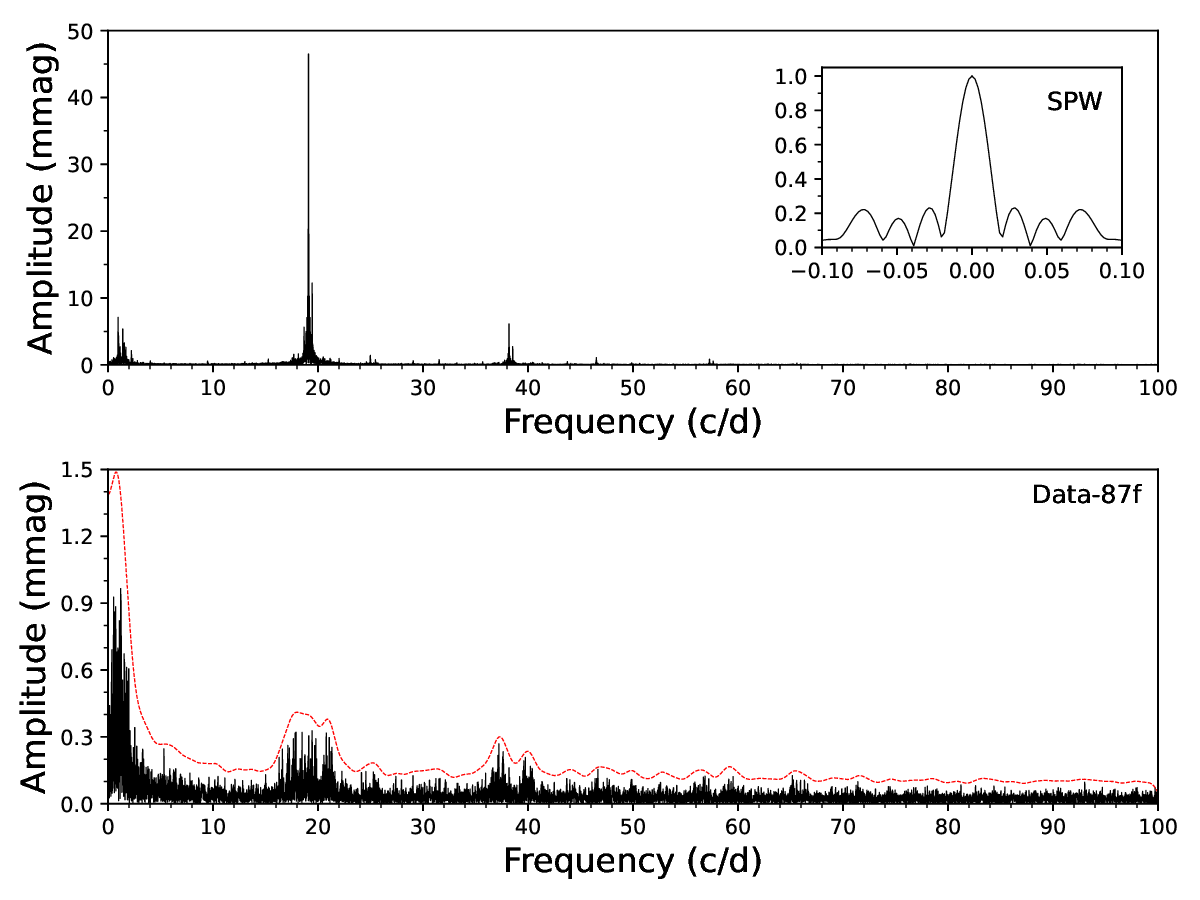}{0.5\textwidth}{(b) TIC 139729335}}         
  \caption{\label{Figure 2} Fourier amplitude spectrum. Panel (a) is for TIC 65138566 and panel (b) is for TIC 139729335. In each panel, the upper subpanel presents the original spectrum where the inset panel shows the window spectrum, and the lower subpanel presents the residual spectrum after the frequencies with S/N $>$ 4 were subtracted. The dashed lines in red show the level of S/N=4.}
\end{figure*}

\begin{figure*}
\gridline{\fig{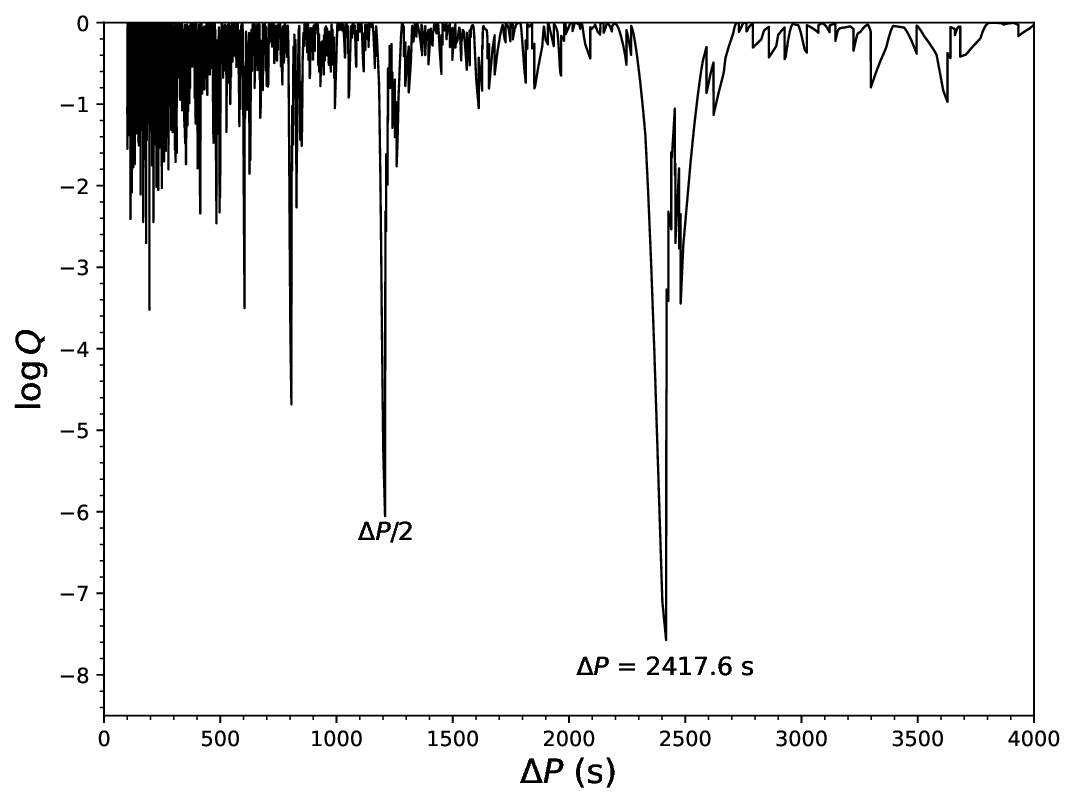}{0.5\textwidth}{(a) TIC 65138566}
          \fig{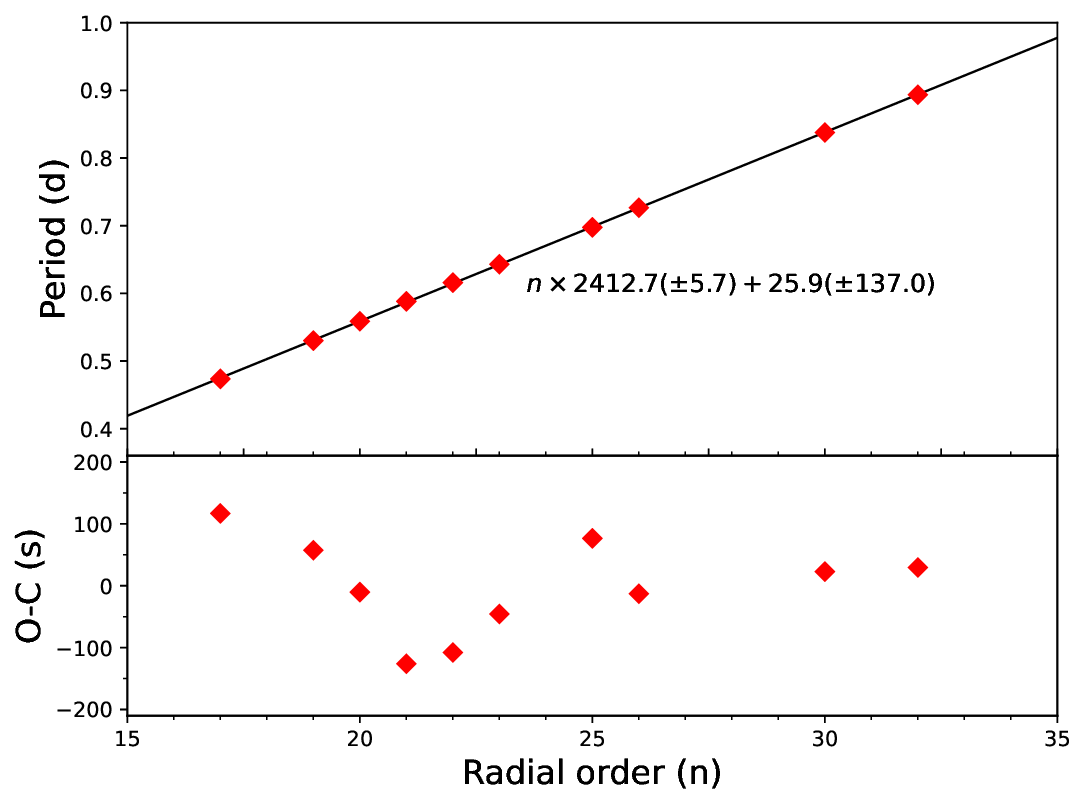}{0.5\textwidth}{(b) TIC 65138566}}
\gridline{\fig{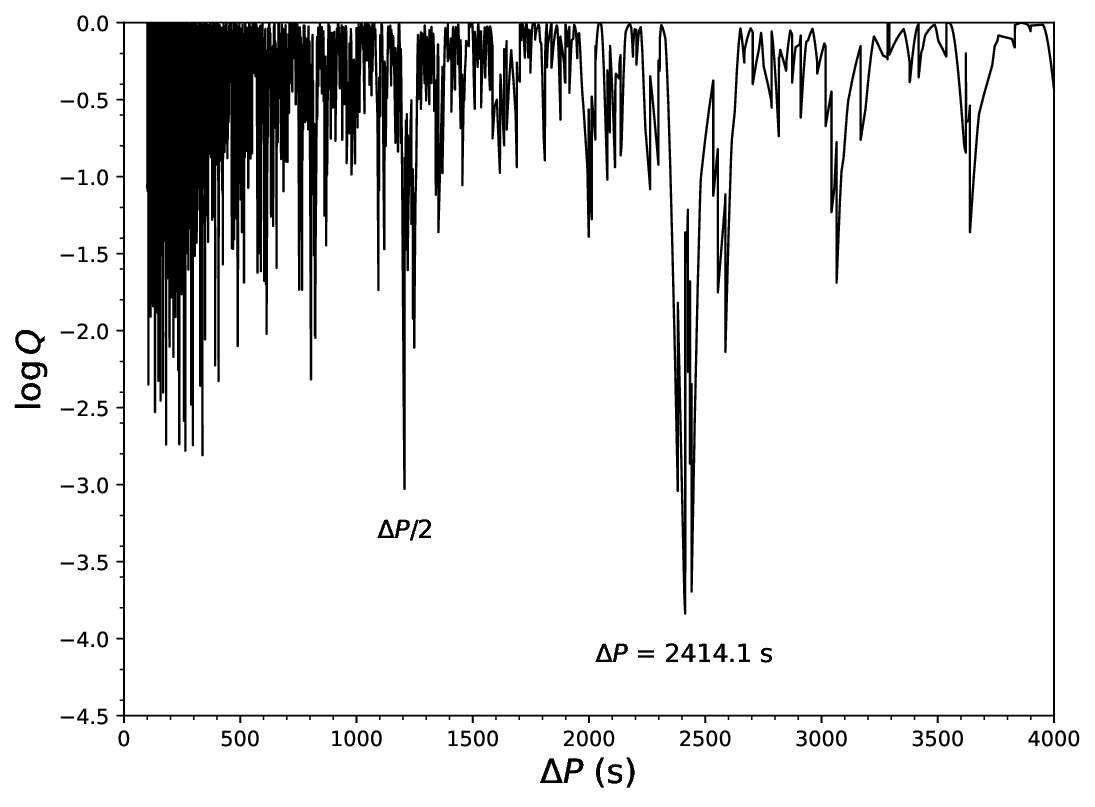}{0.5\textwidth}{(c) TIC 139729335}
          \fig{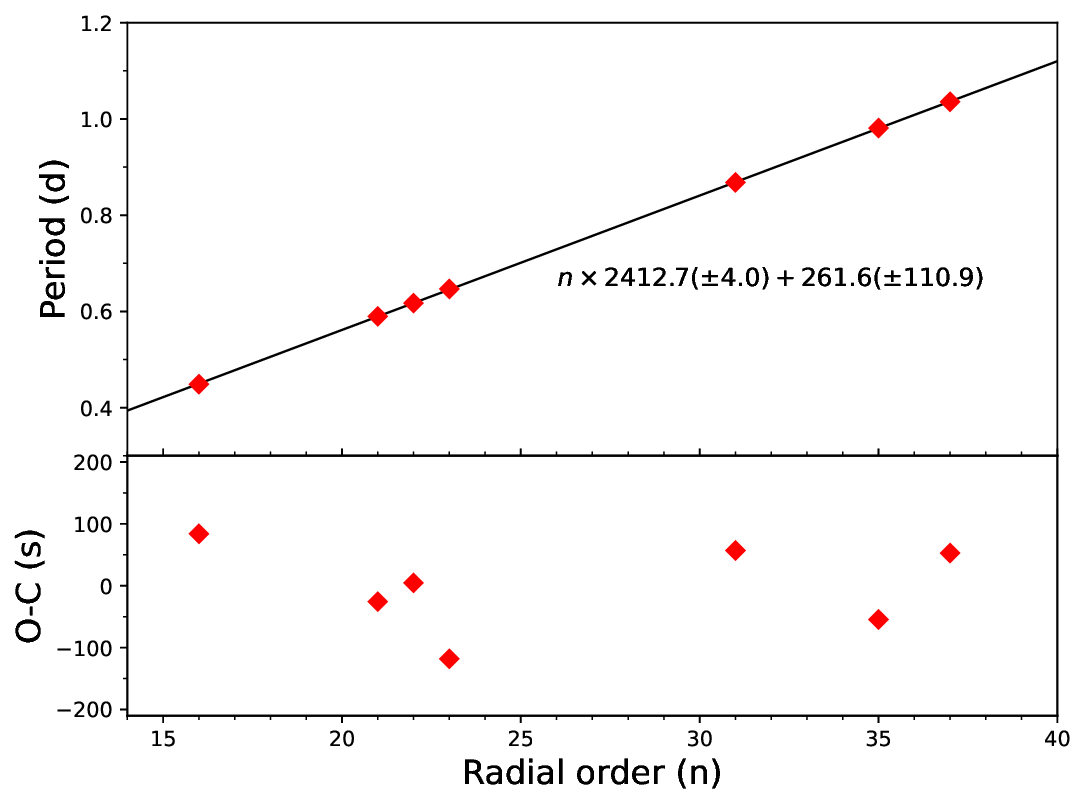}{0.5\textwidth}{(d) TIC 139729335}}
  \caption{\label{Figure 3}  The left panels display the results of the KS test applied to periods of g modes. The right panels show the linear fittings to the $P_n$ .vs. $n$. The panels (a) and (b) are for TIC 65138566, and the panels (c) and (d) are for TIC 139729335.}
\end{figure*}

\begin{figure*}
\gridline{\fig{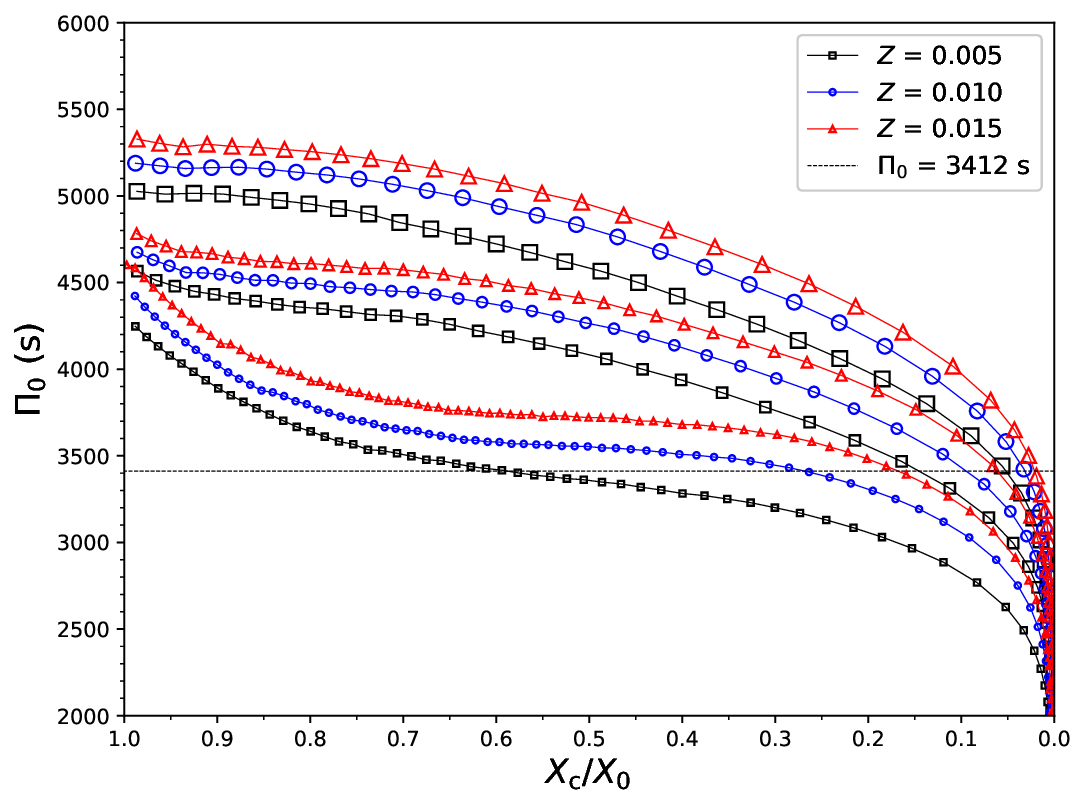}{0.5\textwidth}{(a)}}
\gridline{\fig{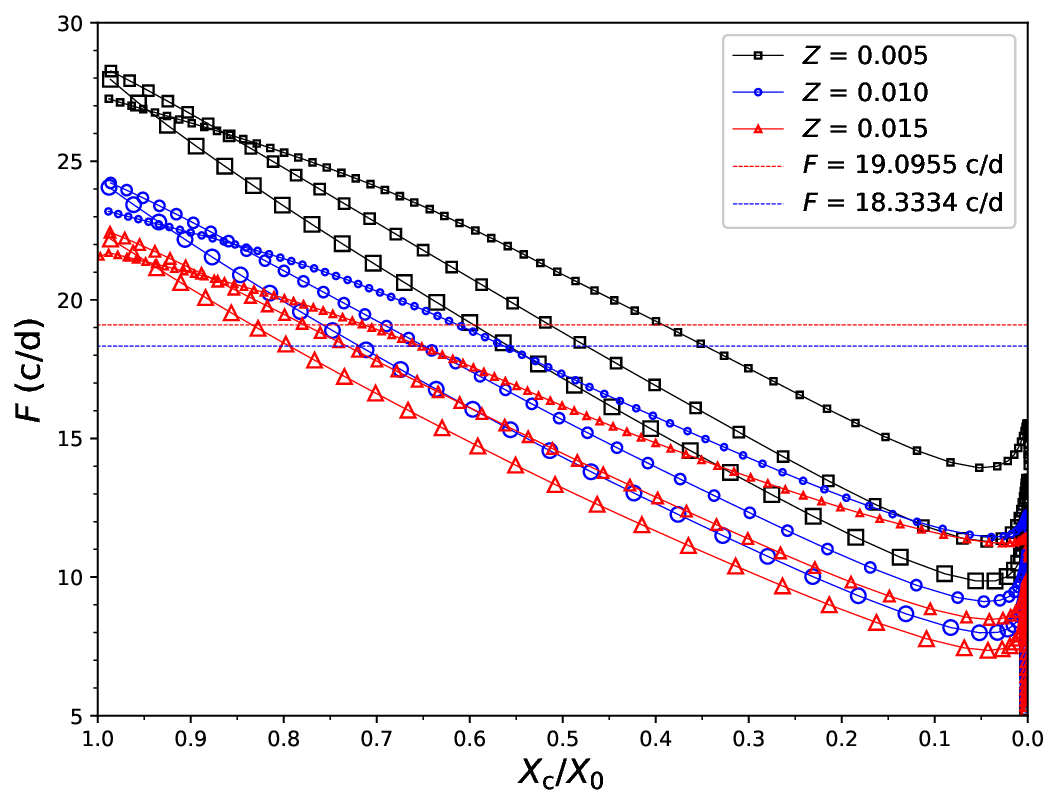}{0.5\textwidth}{(b)}}
  \caption{\label{Figure 4} Evolution of the asymptotic period spacing $\Pi_0$ and the radial fundamental mode $F$ from the zero-age main sequence ($X_{\rm c}/X_0$ = 1) to the terminal-age main sequence($X_{\rm c}/X_0$=0).  $X_{\rm c}$ is the mass fraction of hydrogen in the center of the star. $X_0$ is the initial mass fraction of hydrogen. The black lines with open squares,  blue lines with open circles,  and red lines with open triangles correspond to initial metallicity $Z$ = 0.005, 0.010, and 0.015, respectively. The sizes of the squares, circles, and triangles mark different initial mass $M$ of the star,  smallest size to biggest for 1.30$M_{\odot}$, 1.60$M_{\odot}$, and 1.90$M_{\odot}$, respectively.  }
\end{figure*}

\begin{figure*}
\gridline{\fig{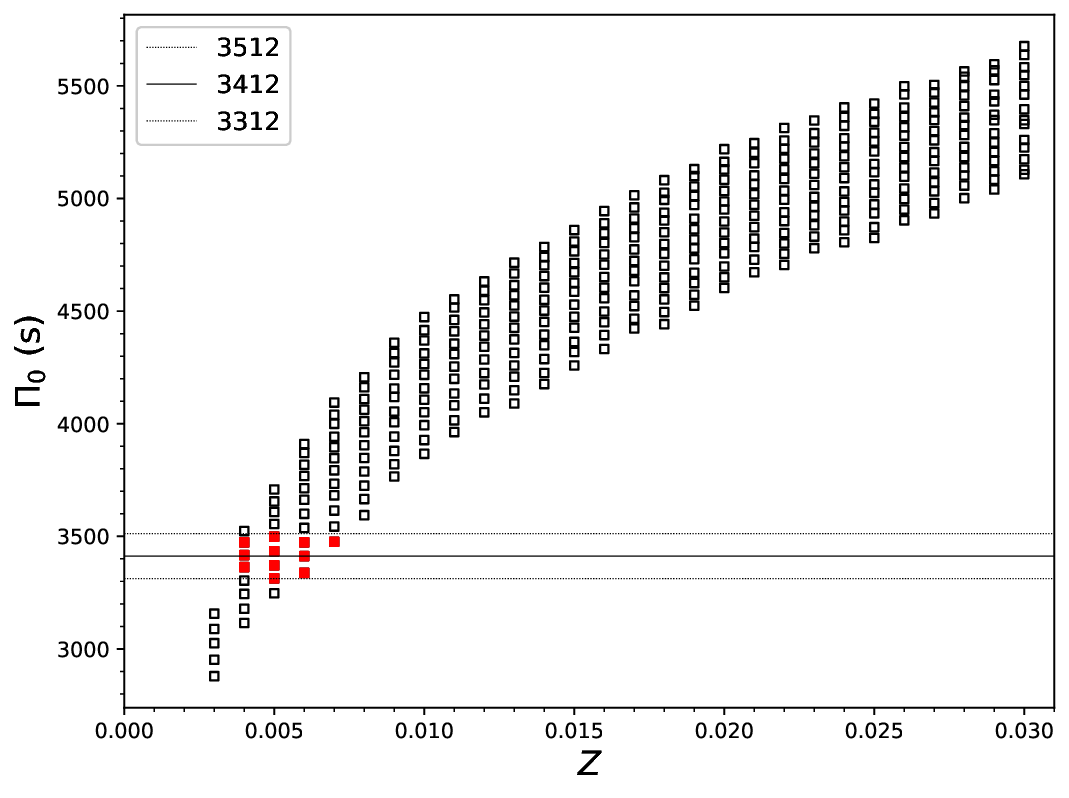}{0.33\textwidth}{(a) TIC 65138566}
          \fig{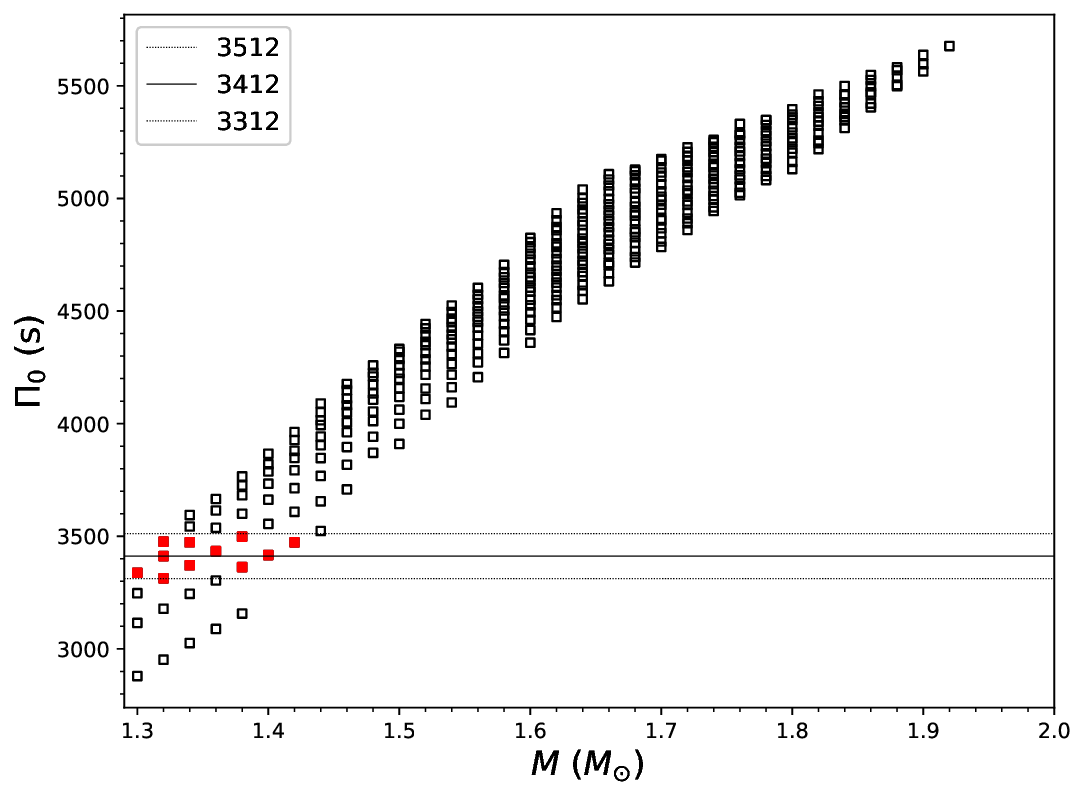}{0.33\textwidth}{(b) TIC 65138566}
          \fig{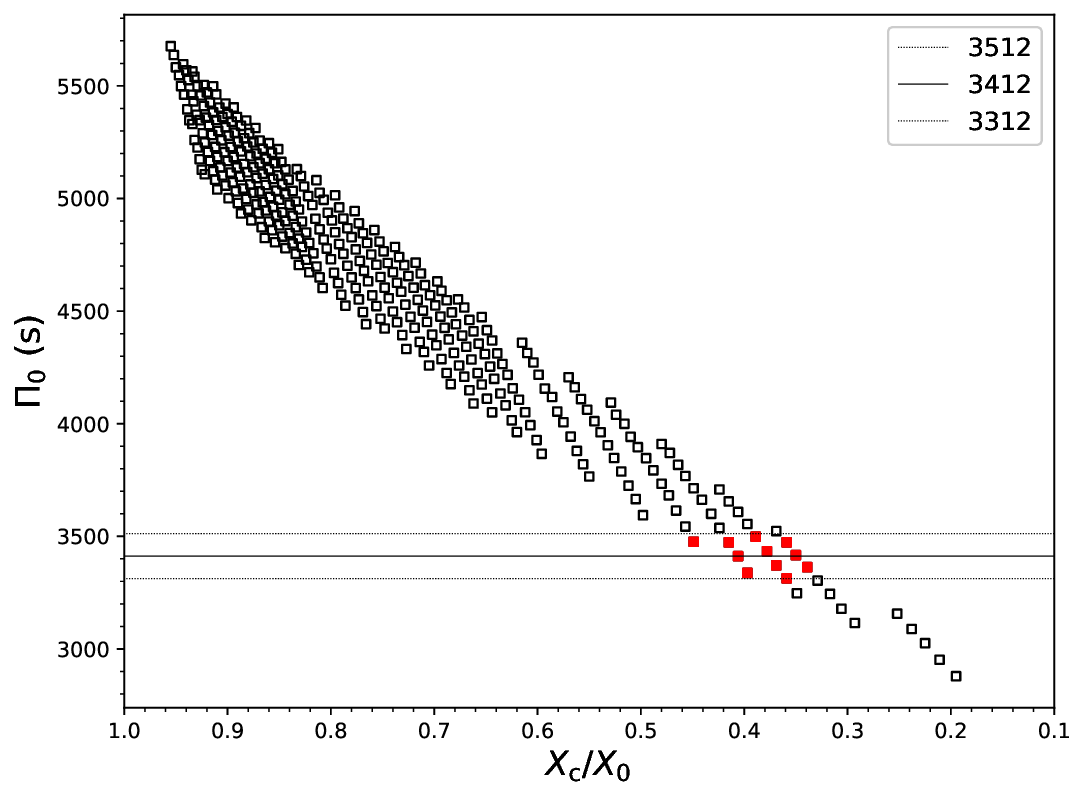}{0.33\textwidth}{(c) TIC 65138566}}
\gridline{\fig{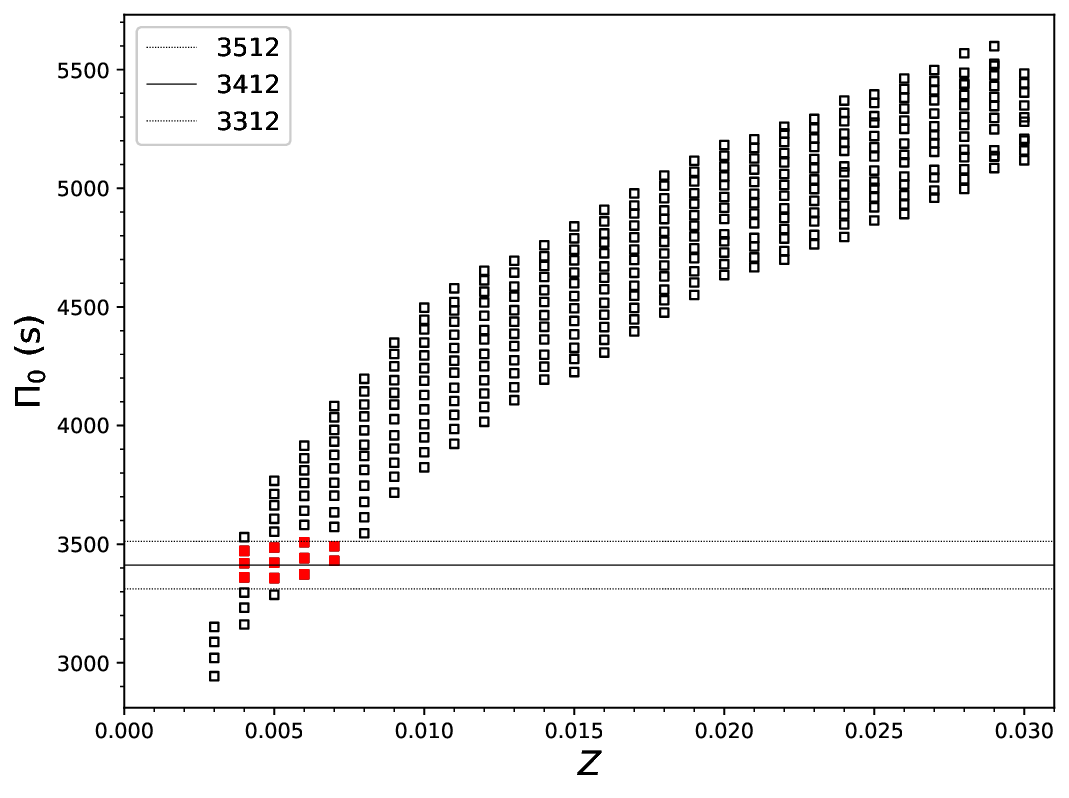}{0.33\textwidth}{(d) TIC 139729335}
          \fig{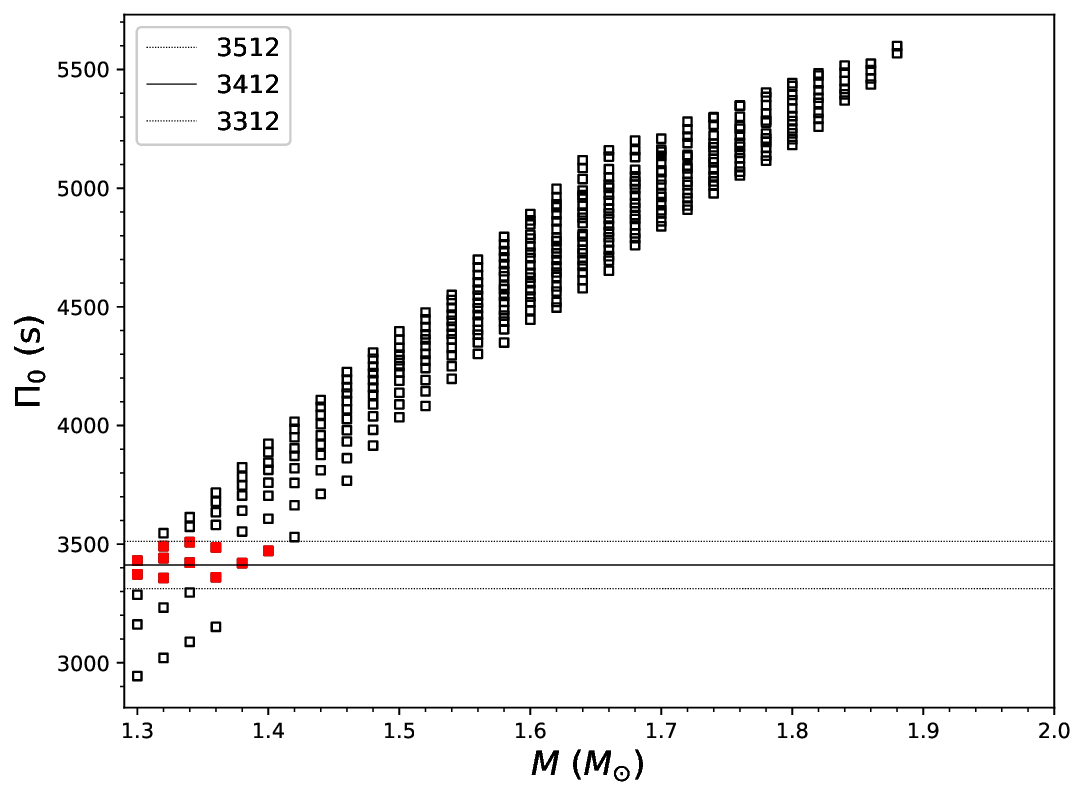}{0.33\textwidth}{(e) TIC 139729335}
          \fig{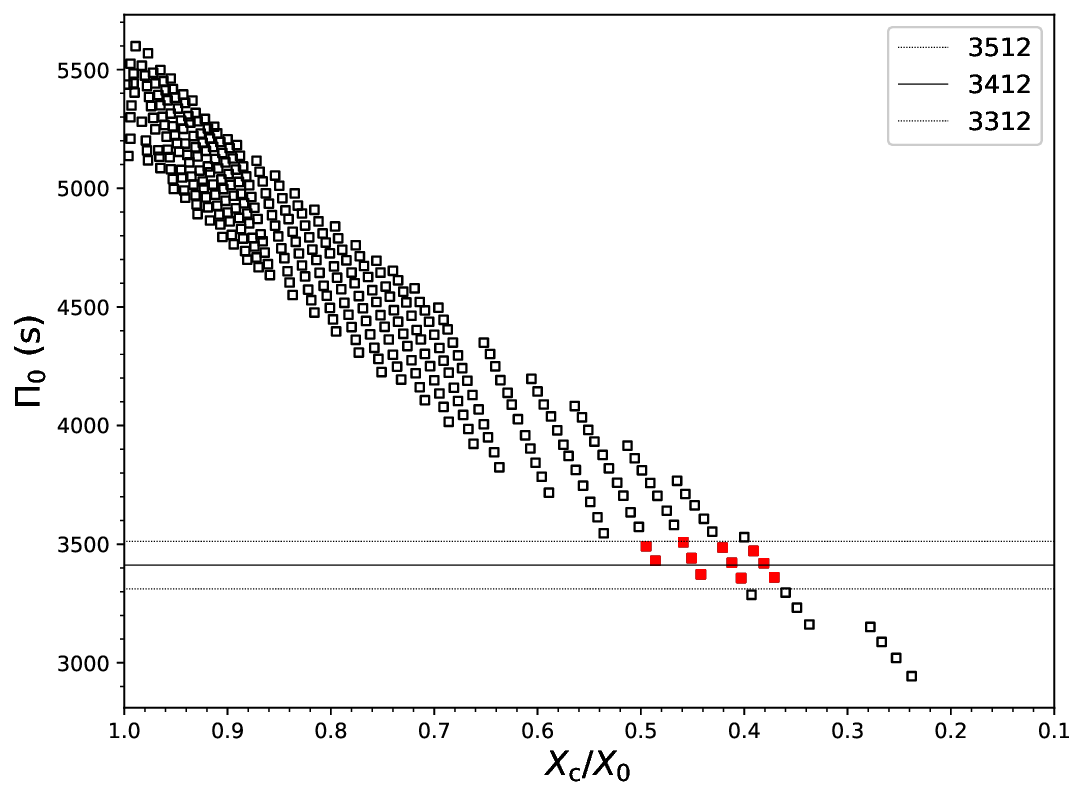}{0.33\textwidth}{(f) TIC 139729335}}
  \caption{\label{Figure 5}  Plots of $\Pi_0$ versus $Z$,  $M$,  and $X_{\rm c}/X_0$,  respectively. Panels (a)-(c) are for TIC 65138566 and panels (d)-(f) are for TIC 139729335. The horizontal lines mark the range of the period spacing 3412 $\pm$ 100 s. Each square in the panels represents the best-fitting model to the radial fundamental mode $F$ along one evolutionary track. The red squares mark the preferred models that fall inside the observed range of the period spacing. }
\end{figure*}
\begin{figure*}
\gridline{\fig{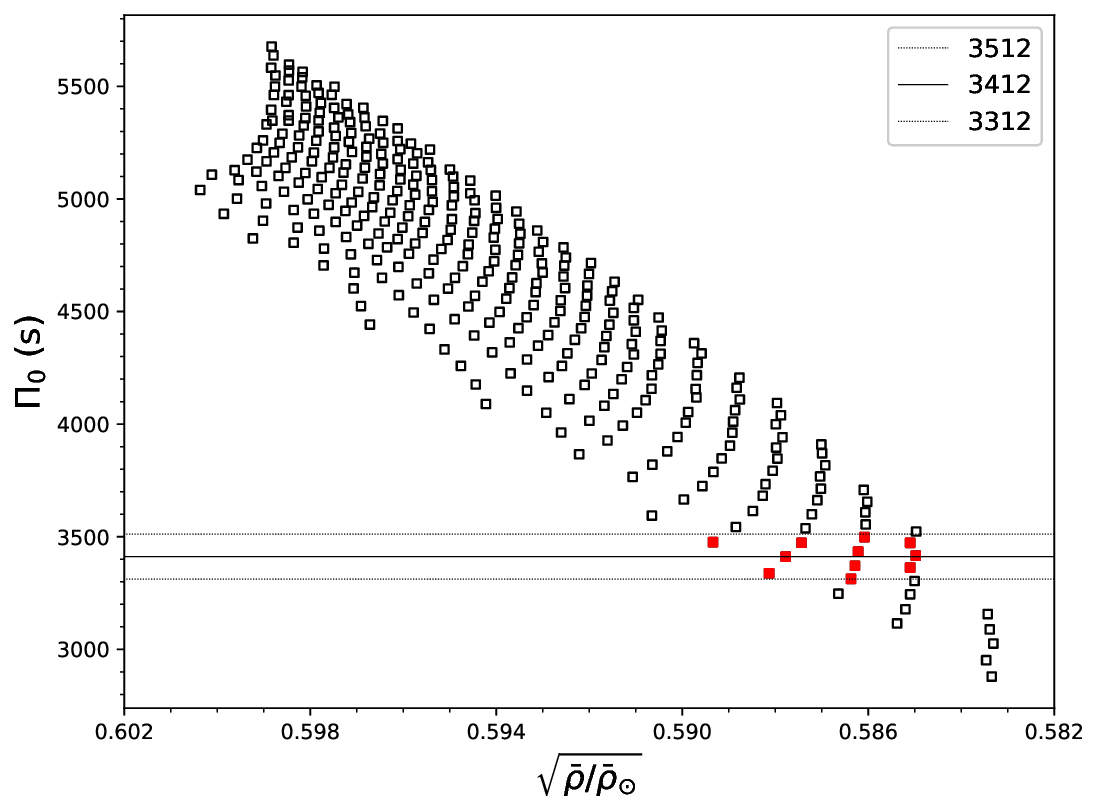}{0.5\textwidth}{(a) TIC 65138566}
          \fig{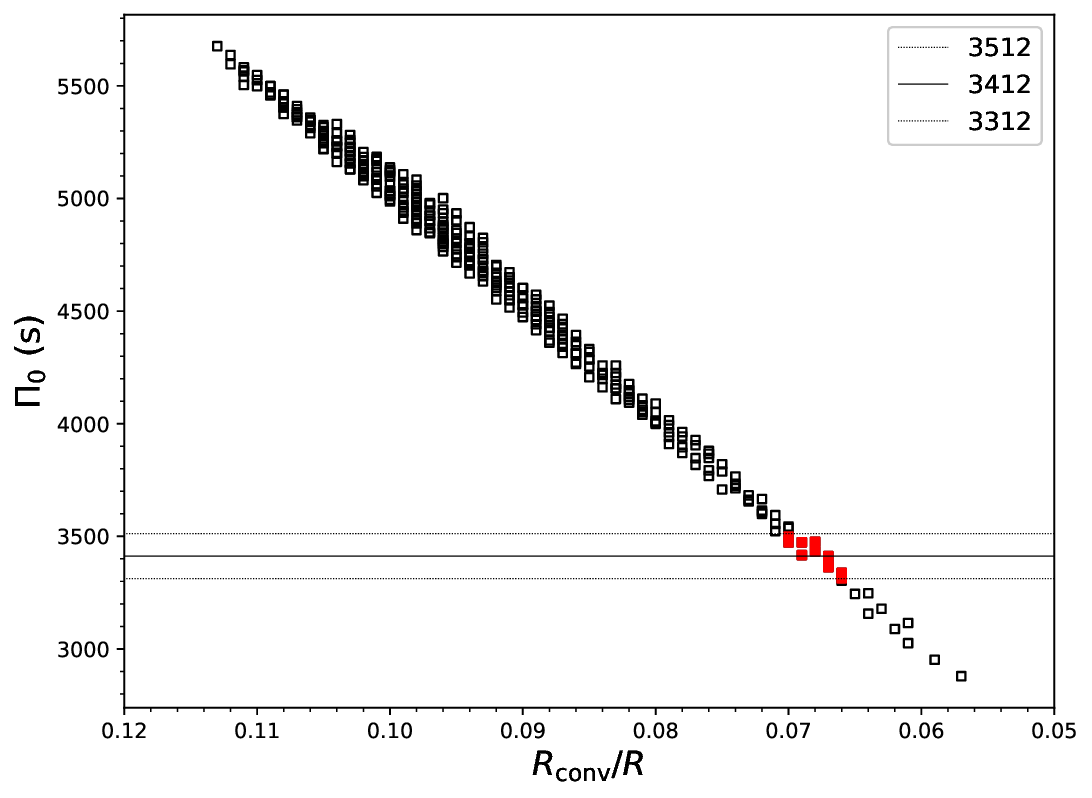}{0.5\textwidth}{(b) TIC 65138566}}
\gridline{\fig{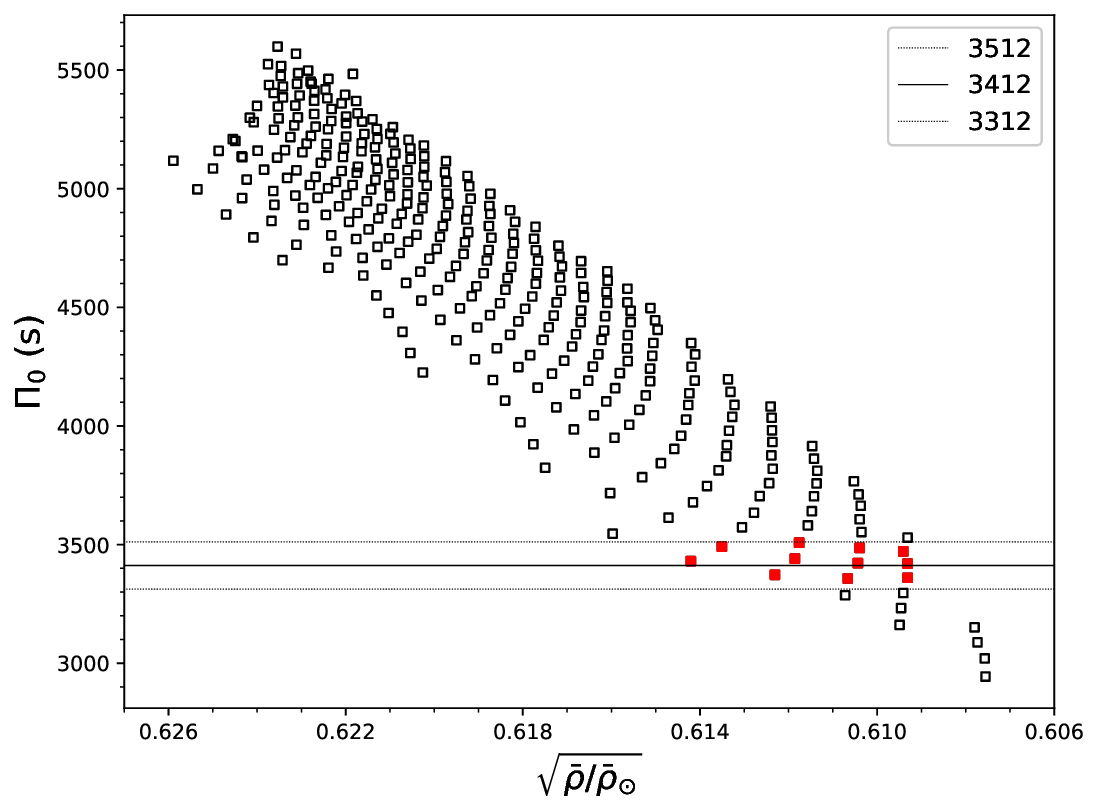}{0.5\textwidth}{(c) TIC 139729335}
          \fig{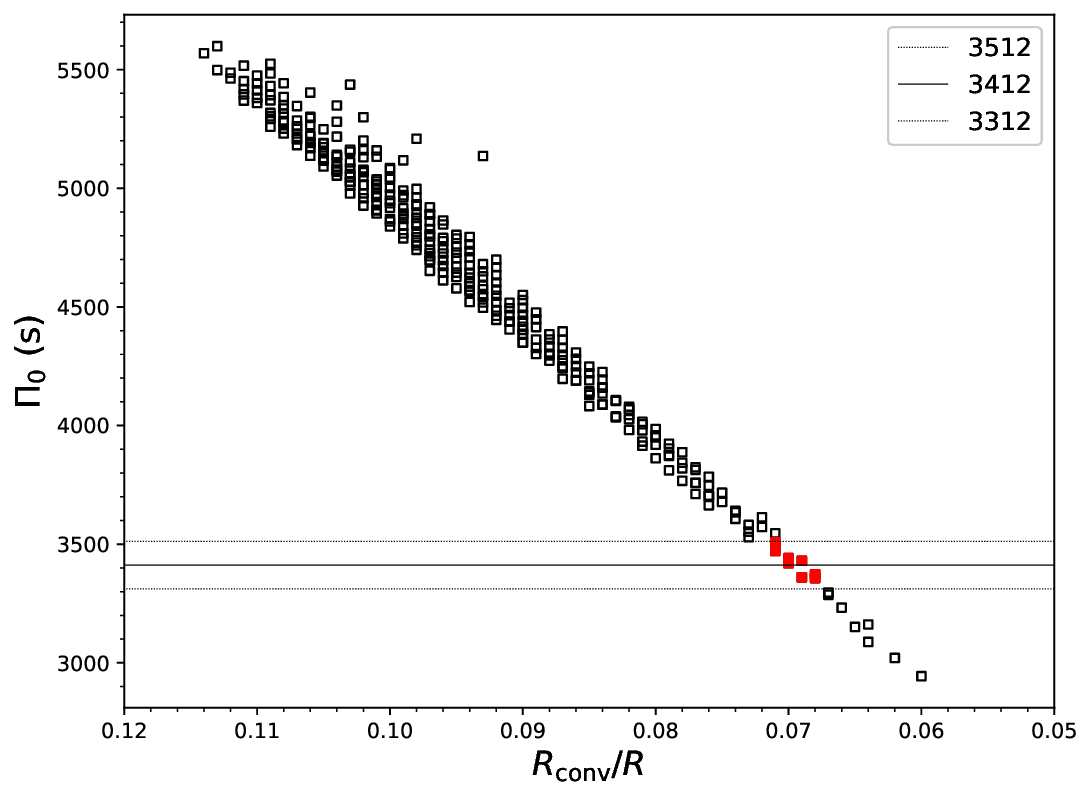}{0.5\textwidth}{(d) TIC 139729335}}                                                
  \caption{\label{Figure 6} Plots of $\Pi_0$ versus the mean density $\sqrt{\bar{\rho}/\bar{\rho}_{\odot}}$ and the relative radius of the convective core $R_{\rm conv}/R$. Panels (a) and (b) are for TIC 65138566, and panels (c) and (d) is for TIC 139729335. Symbols and lines are the same as those in Figure 5.}
\end{figure*}

\begin{table*}
\centering
\footnotesize
\caption{\label{t2} Frequencies detected for TIC 65138566.  }
\begin{tabular}{ccccccccccccccccc}
\hline\hline
 ID    &Freq.                        &Ampl.               &S/N   &Remark     &ID    &Freq.                            &Ampl.               &S/N    &Remark \\
        &(d$^{-1}$)   &($\mu$mag)     &         &                 &        &(d$^{-1}$)     &($\mu$mag)   &        & \\
\hline
$f_{ 1}$ &18.3334(  1) &35506( 37) & 27.70 &$F$            &$f_{37}$ &55.3928( 41) &  560( 23) & 10.68 &$2f_1+f_2$              \\
$f_{ 2}$ &18.7261(  1) &29928( 24) &158.59 &$p_1$          &$f_{38}$ &19.7672( 40) &  556( 23) &  4.95 &$f_1+f_4$              \\
$f_{ 3}$ &23.6429(  4) & 5694( 23) & 36.19 &$p_2$          &$f_{39}$ &19.1184( 40) &  555( 23) &  5.41 &$2f_2-f_1$              \\
$f_{ 4}$ & 1.4337(  4) & 5083( 24) &  8.97 &$g_1$          &$f_{40}$ &24.7175( 43) &  512( 23) & 11.97 &$f_{12}+f_{17}$         \\
$f_{ 5}$ &37.0595(  6) & 4902( 23) & 23.01 &$f_1+f_2$      &$f_{41}$ &20.1593( 49) &  483( 23) &  4.84 &$f_2+f_4$              \\
$f_{ 6}$ & 1.1192(  6) & 3947( 24) &  8.92 &$g_2$          &$f_{42}$ &18.7408( 50) &  471( 23) &  5.24 &Unresolved peak        \\
$f_{ 7}$ &36.6669(  7) & 3179( 23) & 32.20 &$2f_1$         &$f_{43}$ &20.7174( 51) &  453( 23) &  4.63 &$f_{33}-f_{1}$         \\
$f_{ 8}$ & 1.5550(  8) & 2985( 26) &  7.04 &$g_3$          &$f_{44}$ &55.7842( 50) &  450( 23) & 12.44 &$f_1+2f_2$              \\
$f_{ 9}$ &18.3961( 10) & 2630( 25) & 20.27 &$p_3$          &$f_{45}$ &19.4543( 51) &  446( 24) &  4.96 &$f_1+f_6$              \\
$f_{10}$ & 1.7002( 11) & 2186( 24) &  6.18 &$g_4$          &$f_{46}$ &20.3521( 53) &  444( 24) &  5.01 &$f_2+f_{20}$           \\
$f_{11}$ & 1.3765( 11) & 1971( 23) &  6.46 &$g_5$          &$f_{47}$ &30.3728( 62) &  394( 24) & 12.34 &$p_6$              \\
$f_{12}$ & 1.2640( 13) & 1936( 24) &  7.44 &$g_6$          &$f_{48}$ &55.0003( 62) &  355( 22) &  9.69 &$3f_1$              \\
$f_{13}$ &42.3689( 12) & 1873( 23) & 26.67 &$f_2+f_3$      &$f_{49}$ &34.3729( 64) &  345( 23) & 11.87 &$f_{30}-f_{24}$         \\
$f_{14}$ & 0.3919( 13) & 1774( 24) &  7.68 &$f_2-f_1$      &$f_{50}$ &36.7477( 67) &  345( 24) &  7.48 &$f_{1}+f_{9}$           \\
$f_{15}$ & 1.7892( 13) & 1749( 23) &  5.91 &$g_7$          &$f_{51}$ &20.5156( 67) &  343( 23) &  4.55 &$f_2+f_{15}$            \\
$f_{16}$ & 1.1937( 14) & 1641( 24) &  6.98 &$g_8$          &$f_{52}$ &13.4165( 67) &  337( 22) &  9.19 &$f_5-f_3$              \\
$f_{17}$ &23.4521( 22) & 1516( 29) & 19.72 &$p_4$          &$f_{53}$ &20.4276( 67) &  327( 23) &  4.49 &$f_2+f_{10}$            \\
$f_{18}$ &37.4517( 15) & 1471( 23) & 22.20 &$2f_2$         &$f_{54}$ &60.7017( 72) &  319( 24) &  9.80 &$f_3+f_5$              \\
$f_{19}$ & 5.3097( 18) & 1213( 23) & 13.16 &$f_3-f_1$      &$f_{55}$ &21.2476( 76) &  300( 23) &  4.77 &$f_{22}-f_{43}$          \\
$f_{20}$ & 1.6237( 25) &  942( 24) &  4.99 &$g_9$          &$f_{56}$ &35.6255( 77) &  291( 22) &  5.93 &$f_5-f_{4}$              \\
$f_{21}$ &24.2081( 24) &  939( 22) & 16.36 &$p_5$          &$f_{57}$ &29.0687( 91) &  255( 22) &  8.96 &$f_{49}-f_{19}$          \\
$f_{22}$ &41.9760( 24) &  913( 23) & 21.36 &$f_1+f_3$      &$f_{58}$ &41.7860( 98) &  235( 23) &  6.36 &$f_1+f_{17}$              \\
$f_{23}$ &17.2907( 25) &  903( 23) &  5.94 &$f_2-f_4$      &$f_{59}$ &35.9421(104) &  221( 23) &  4.97 &$f_5-f_6$              \\
$f_{24}$ & 1.8861( 25) &  864( 22) &  4.53 &$g_{10}$       &$f_{60}$ &37.0793(171) &  206( 24) &  5.40 &Unresolved peak           \\
$f_{25}$ &16.7782( 28) &  851( 23) &  6.04 &$f_1-f_8$      &$f_{61}$ &23.2525(113) &  203( 23) &  4.57 &$f_1+f_{29}$              \\
$f_{26}$ &17.6061( 27) &  828( 22) &  5.65 &$f_2-f_6$      &$f_{62}$ &61.0947(118) &  198( 23) &  7.69 &$2f_2+f_3$              \\
$f_{27}$ &16.9023( 29) &  790( 24) &  6.02 &$f_1-f_4$      &$f_{63}$ &55.4540(118) &  195( 23) &  6.16 &$f_5+f_9$              \\
$f_{28}$ &37.1222( 30) &  786( 23) & 14.64 &$f_2+f_9$      &$f_{64}$ &23.4422(139) &  192( 28) &  4.63 &Unresolved peak          \\
$f_{29}$ & 4.9175( 30) &  758( 24) &  9.83 &$f_3-f_2$      &$f_{65}$ &60.3093(121) &  189( 22) &  6.85 &$2f_1+f_3$              \\
$f_{30}$ &36.2674( 33) &  671( 23) & 11.70 &$2f_1-f_{14}$  &$f_{66}$ &24.0377(120) &  187( 22) &  4.41 &$f_{13}-f_1$              \\
$f_{31}$ & 2.1112( 35) &  665( 23) &  4.62 &$g_{11}$       &$f_{67}$ &25.2814(126) &  186( 24) &  5.15 &$f_3+f_{20}$              \\
$f_{32}$ &18.3405( 33) &  651( 37) &  5.57 &Unresolved peak&$f_{68}$ &30.3418(141) &  183( 23) &  6.61 &Unresolved peak        \\
$f_{33}$ &39.0504( 37) &  612( 24) & 11.21 &$2f_2+f_{20}$  &$f_{69}$ &56.1793(150) &  155( 23) &  5.94 &$3f_2$              \\
$f_{34}$ &23.4756( 50) &  596( 23) & 11.57 &Unresolved peak&$f_{70}$ &74.1198(183) &  136( 23) &  4.90 &$2f_5$              \\
$f_{35}$ &20.2815( 41) &  584( 24) &  4.85 &$f_2+f_8$      &$f_{71}$ &43.4462(179) &  134( 23) &  5.25 &$f_2+f_{40}$           \\
$f_{36}$ &17.2192( 40) &  569( 23) &  4.63 &$f_1-f_6$      &$f_{72}$ &74.5093(235) &  111( 23) &  4.70 &$2f_2+f_5$            \\
\hline
\end{tabular}
\end{table*}
\newpage

\begin{table*}
\centering
\footnotesize
\caption{\label{t2} Frequencies detected for TIC 139729335. }
\begin{tabular}{ccccccccccccccccc}
\hline\hline
 ID    &Freq.                        &Ampl.               &S/N   &Remark     &ID    &Freq.                            &Ampl.               &S/N    &Remark \\
        &(d$^{-1}$)   &($\mu$mag)     &         &                 &        &(d$^{-1}$)     &($\mu$mag)   &        & \\
\hline
$f_{ 1}$ &19.0955(  1) &46525( 84) & 60.87 &$F$                 &$f_{45}$ &17.9447( 67) &  478( 31) &  4.44 &$f_{1}-f_{10}$ \\
$f_{ 2}$ &19.4464(  3) &11891( 30) & 52.58 &$p_1$               &$f_{46}$ &35.6783( 75) &  450( 30) &  9.70 &$f_{15}+f_{29}$\\
$f_{ 3}$ & 0.9656(  5) & 7099( 31) & 10.60 &$g_1$               &$f_{47}$ &20.2478( 74) &  438( 32) &  4.50 &$f_{1}+f_{10}$ \\
$f_{ 4}$ &38.1913(  5) & 6132( 32) & 31.63 &2$f_1$              &$f_{48}$ &17.4761( 83) &  433( 33) &  4.46 &$f_{1}-f_{13}$ \\
$f_{ 5}$ & 1.4031(  6) & 5402( 32) &  9.02 &$g_2$               &$f_{49}$ &21.1353( 75) &  418( 32) &  4.38 &Unresolved peak \\
$f_{ 6}$ &18.6752(  6) & 5134( 31) & 31.72 &$p_2$               &$f_{50}$ &38.1227( 81) &  416( 31) &  6.55 &$f_{2}+f_{6}$  \\
$f_{ 7}$ & 1.0193( 11) & 3208( 31) &  6.57 &$g_3$               &$f_{51}$ &37.6170( 80) &  406( 31) &  5.57 &$f_{18}+f_{29}$\\
$f_{ 8}$ & 1.5464( 12) & 2863( 33) &  6.45 &$g_4$               &$f_{52}$ &38.8930( 86) &  380( 30) &  6.98 &2$f_{2}$       \\
$f_{ 9}$ &38.5421( 11) & 2738( 30) & 36.33 &$f_1+f_2$           &$f_{53}$ &20.4140( 85) &  380( 32) &  4.28 &$f_{2}+f_{3}$  \\
$f_{10}$ & 1.1520( 13) & 2436( 30) &  6.34 &$g_5$               &$f_{54}$ &40.4364( 83) &  369( 30) &  5.45 &2$f_{2}+f_{8}$ \\
$f_{11}$ & 1.6957( 14) & 2354( 35) &  6.73 &$g_6$               &$f_{55}$ &33.2170(139) &  344( 33) & 10.25 &$f_{21}+f_{45}$ \\
$f_{12}$ & 2.2279( 16) & 2100( 31) &  6.82 &$g_7$               &$f_{56}$ &41.3523( 94) &  341( 31) &  7.52 &2$f_{5}+f_{9}$  \\
$f_{13}$ & 1.6199( 18) & 1926( 34) &  6.67 &$g_8$               &$f_{57}$ &24.6155( 96) &  338( 30) &  7.37 &$f_{51}-f_{38}$ \\
$f_{14}$ &24.9826( 24) & 1360( 29) & 17.88 &$p_3$               &$f_{58}$ &40.5197(100) &  321( 33) &  5.51 &2$f_{2}+f_{13}$ \\
$f_{15}$ &18.1306( 28) & 1264( 31) &  7.10 &$f_1-f_3$           &$f_{59}$ &46.4947(103) &  320( 33) &  6.43 &Unresolved peak \\
$f_{16}$ &17.6921( 24) & 1244( 32) &  7.63 &$f_1-f_5$           &$f_{60}$ &40.2458(100) &  311( 31) &  5.34 &$f_{1}+f_{17}$  \\
$f_{17}$ &21.1503( 28) & 1119( 33) &  8.13 &$f_2+f_{11}$        &$f_{61}$ &49.8617(108) &  287( 31) &  7.07 &$p_{10}$        \\
$f_{18}$ &20.0592( 33) & 1088( 31) &  7.69 &$f_1+f_3$           &$f_{62}$ &36.7872(122) &  274( 32) &  4.14 &2$f_{1}-f_{5}$  \\
$f_{19}$ &46.5106( 32) & 1035( 33) & 17.66 &$p_4$               &$f_{63}$ &29.0886(171) &  272( 30) &  6.92 &Unresolved peak \\
$f_{20}$ &22.0108( 33) &  936( 31) & 12.93 &$p_5$               &$f_{64}$ &46.4684(154) &  251( 33) &  5.66 &Unresolved peak \\
$f_{21}$ &15.2687( 35) &  912( 31) & 21.80 &$p_6$               &$f_{65}$ &65.6069(127) &  243( 31) &  6.69 &$f_{1}+f_{19}$  \\
$f_{22}$ &20.5006( 37) &  889( 31) &  6.49 &$f_1+f_5$           &$f_{66}$ &34.9044(161) &  223( 32) &  5.93 &$f_{11}+f_{55}$ \\
$f_{23}$ &25.4672( 37) &  871( 30) & 17.38 &$p_7$               &$f_{67}$ &44.5618(164) &  211( 30) &  5.40 &$f_{1}+f_{23}$  \\
$f_{24}$ &57.2868( 38) &  850( 30) & 16.48 &3$f_1$              &$f_{68}$ &41.1055(158) &  205( 31) &  5.28 &$f_{1}+f_{20}$  \\
$f_{25}$ & 2.3677( 37) &  832( 31) &  5.02 &$f_3+f_5$           &$f_{69}$ &39.1579(175) &  200( 30) &  4.17 &2$f_{1}+f_{3}$  \\
$f_{26}$ &31.5374( 41) &  759( 30) & 18.61 &$p_8$               &$f_{70}$ &26.4013(167) &  196( 30) &  6.01 &$f_{5}+f_{14}$  \\
$f_{27}$ &37.7696( 45) &  704( 33) &  7.95 &$f_1+f_6$           &$f_{71}$ &47.2371(178) &  196( 32) &  4.68 &$f_{15}+f_{63}$ \\
$f_{28}$ &29.0634( 67) &  668( 33) & 14.29 &$p_9$               &$f_{72}$ &44.0803(185) &  193( 29) &  4.76 &$f_{1}+f_{14}$  \\
$f_{29}$ &17.5498( 55) &  624( 33) &  4.44 &$f_1-f_8$           &$f_{73}$ &27.9166(202) &  167( 30) &  4.79 &$f_{28}-f_{10}$ \\
$f_{30}$ & 9.4850( 52) &  604( 31) & 12.82 &$f_9-f_{28}$        &$f_{74}$ &43.7717(245) &  167( 32) &  4.45 &Unresolved peak \\
$f_{31}$ &18.0778( 59) &  596( 33) &  4.52 &$f_1-f_7$           &$f_{75}$ &30.2093(212) &  166( 31) &  4.46 &$f_{10}+f_{28}$ \\
$f_{32}$ &57.6380( 53) &  591( 31) & 17.84 &2$f_1+f_2$          &$f_{76}$ &50.6303(216) &  166( 31) &  5.30 &$f_{1}+f_{26}$  \\
$f_{33}$ & 4.0286( 55) &  569( 31) &  6.74 &$g_9$               &$f_{77}$ &24.0449(223) &  163( 30) &  4.28 &$f_{14}-f_{3}$  \\
$f_{34}$ &20.6439( 58) &  538( 31) &  4.44 &$f_1+f_8$           &$f_{78}$ &28.5798(231) &  161( 30) &  4.78 &$f_{1}+f_{30}$  \\
$f_{35}$ & 2.7926( 59) &  529( 30) &  4.50 &2$f_5$              &$f_{79}$ &34.3637(221) &  160( 30) &  4.78 &$f_{1}+f_{21}$  \\
$f_{36}$ &19.0705( 88) &  520( 34) &  4.47 &Unresolved peak     &$f_{80}$ &52.5641(234) &  155( 30) &  4.48 &$f_{26}+f_{41}$ \\
$f_{37}$ &16.8669( 64) &  512( 31) &  5.20 &$f_1-f_{12}$        &$f_{81}$ &49.8218(241) &  154( 31) &  4.15 &$f_{38}+f_{62}$ \\
$f_{38}$ &13.0303( 62) &  511( 30) & 12.55 &$f_{21}-f_{12}$     &$f_{82}$ &57.9891(287) &  141( 32) &  4.69 &$f_{1}+2f_{2}$  \\
$f_{39}$ &43.7472( 69) &  499( 33) & 10.59 &$f_{19}-f_{35}$     &$f_{83}$ &76.3838(306) &  130( 31) &  5.03 &4$f_{1}$        \\
$f_{40}$ &20.1116( 72) &  495( 31) &  4.41 &$f_{1}+f_{7}$       &$f_{84}$ &33.1898(386) &  129( 31) &  4.30 &Unresolved peak \\
$f_{41}$ &21.0047( 66) &  493( 31) &  4.45 &$f_{2}+f_{8}$       &$f_{85}$ &54.7790(295) &  121( 30) &  4.43 &$f_{1}+f_{46}$  \\
$f_{42}$ &18.0423( 67) &  491( 31) &  4.23 &$f_{2}-f_{5}$       &$f_{86}$ &63.1768(329) &  120( 31) &  4.10 &2$f_{1}+f_{14}$ \\
$f_{43}$ &17.3998( 70) &  489( 33) &  4.33 &$f_{1}-f_{11}$      &$f_{87}$ &62.8424(337) &  118( 31) &  4.30 &$f_{1}+f_{39}$  \\
$f_{44}$ &19.0913( 99) &  487( 83) &  4.59 &Unresolved peak\\
\hline
\end{tabular}
\end{table*}
\newpage

\begin{table*}
\centering
\footnotesize
\caption{\label{t5} Regular low-frequency g modes. }
\begin{tabular}{ccccccccccccccccc}
\hline\hline
TIC &ID   &Frequency        &Period           &Radial order   \\
    &     &(d$^{-1}$)       &(s)              &              \\
\hline
           &$f_{31}$  &2.1112 & 40924.1 &  17 \\
           &$f_{24}$  &1.8861 & 45808.9 &  19 \\
           &$f_{15}$  &1.7892 & 48289.2 &  20 \\
           &$f_{10}$  &1.7002 & 50817.7 &  21 \\
65138566   &$f_{20}$  &1.6237 & 53212.1 &  22 \\
           &$f_{8}$   &1.5550 & 55562.4 &  23 \\
           &$f_{4}$   &1.4337 & 60265.6 &  25 \\
           &$f_{11}$  &1.3765 & 62767.8 &  26 \\
           &$f_{16}$  &1.1937 & 72382.6 &  30 \\
           &$f_{6}$  &1.1192 & 77201.1 &  32 \\
\hline
           &$f_{12}$  &2.2279 & 38780.5 &  16 \\
           &$f_{11}$  &1.6957 & 50953.6 &  21 \\
           &$f_{13}$  &1.6199 & 53335.8 &  22 \\
139729335  &$f_{8}$   &1.5464 & 55871.3 &  23 \\
           &$f_{10}$  &1.1520 & 74997.7 &  31 \\
           &$f_{7}$   &1.0193 & 84759.9 &  35 \\
           &$f_{3}$   &0.9656 & 89477.8 &  37 \\
\hline
\end{tabular}
\end{table*}

\begin{table*}
\centering
\footnotesize
\caption{\label{t1} Preferred models. $\tau_0$ is the acoustic radius.}
\begin{tabular}{ccccccccccccccccc}
\hline\hline 
TIC &ID &$Z$  &$M$           &$f_{\rm ov}$  &$T_{\rm eff}$ &log$g$  &$R$        &$\log (L/L_{\odot})$           &$\tau_0$ &$F$   &$\Pi_0$   &Age   &$X_{\rm c}$  &$X_{\rm C}/X_0$ &$\bar{\rho}$ &$R_{\rm conv}$/R \\
&   &     &($M_{\odot}$) &              &(K)           &(cgs)   &$(R_{\odot})$  &       &(hr)     &(d$^{-1}$)  &(s)       &(Gyr) &             &                &($\bar{\rho}_{\odot}$)  \\
\hline
          &1 &0.004 &1.38 &0.010 & 7723 &4.174 &1.592 &0.908 & 1.84 &18.3359 &  3363 &1.66 &0.252  &0.339  &0.342  &0.067  \\
          &2 &0.004 &1.40 &0.010 & 7812 &4.176 &1.599 &0.932 & 1.84 &18.3322 &  3416 &1.58 &0.259  &0.350  &0.342  &0.069  \\
          &3 &0.004 &1.42 &0.010 & 7900 &4.178 &1.607 &0.956 & 1.84 &18.3358 &  3473 &1.50 &0.266  &0.359  &0.342  &0.070  \\
          &4 &0.005 &1.32 &0.010 & 7319 &4.169 &1.566 &0.801 & 1.82 &18.3320 &  3313 &1.92 &0.265  &0.359  &0.344  &0.066  \\
          &5 &0.005 &1.34 &0.010 & 7407 &4.171 &1.574 &0.826 & 1.82 &18.3342 &  3371 &1.82 &0.272  &0.369  &0.344  &0.067  \\
65138566  &6 &0.005 &1.36 &0.010 & 7494 &4.173 &1.582 &0.851 & 1.82 &18.3350 &  3434 &1.73 &0.279  &0.378  &0.344  &0.068  \\
          &7 &0.005 &1.38 &0.010 & 7582 &4.175 &1.590 &0.875 & 1.83 &18.3330 &  3499 &1.65 &0.287  &0.389  &0.343  &0.070  \\
          &8 &0.006 &1.30 &0.010 & 7104 &4.169 &1.555 &0.743 & 1.80 &18.3322 &  3338 &2.01 &0.292  &0.397  &0.346  &0.066  \\
          &9 &0.006 &1.32 &0.010 & 7190 &4.170 &1.563 &0.768 & 1.81 &18.3345 &  3412 &1.91 &0.299  &0.406  &0.345  &0.067  \\
          &10 &0.006 &1.34 &0.010 & 7275 &4.172 &1.572 &0.794 & 1.81 &18.3322 &  3473 &1.81 &0.305  &0.415  &0.345  &0.069  \\
          &11 &0.007 &1.32 &0.010 & 7066 &4.172 &1.561 &0.737 & 1.79 &18.3322 &  3476 &1.89 &0.329  &0.449  &0.347  &0.068  \\
\hline
          &1 &0.004 &1.36 &0.010 & 7703 &4.196 &1.542 &0.876 & 1.76 &19.0952 &  3360 &1.67 &0.275  &0.371  &0.371  &0.069  \\
          &2 &0.004 &1.38 &0.010 & 7793 &4.198 &1.549 &0.900 & 1.77 &19.0957 &  3420 &1.59 &0.283  &0.381  &0.371  &0.070  \\
          &3 &0.004 &1.40 &0.010 & 7883 &4.200 &1.556 &0.924 & 1.77 &19.0979 &  3472 &1.51 &0.290  &0.391  &0.371  &0.071  \\
          &4 &0.005 &1.32 &0.010 & 7384 &4.192 &1.524 &0.793 & 1.75 &19.0969 &  3357 &1.83 &0.298  &0.403  &0.373  &0.068  \\
          &5 &0.005 &1.34 &0.010 & 7472 &4.194 &1.532 &0.818 & 1.75 &19.0934 &  3422 &1.74 &0.305  &0.412  &0.373  &0.070  \\
139729335 &6 &0.005 &1.36 &0.010 & 7560 &4.197 &1.540 &0.842 & 1.76 &19.0943 &  3486 &1.65 &0.311  &0.421  &0.373  &0.071  \\
          &7 &0.006 &1.30 &0.010 & 7163 &4.192 &1.514 &0.734 & 1.73 &19.0969 &  3373 &1.90 &0.326  &0.442  &0.375  &0.068  \\
          &8 &0.006 &1.32 &0.010 & 7250 &4.194 &1.522 &0.760 & 1.74 &19.0942 &  3441 &1.81 &0.332  &0.451  &0.374  &0.070  \\
          &9 &0.006 &1.34 &0.010 & 7337 &4.196 &1.530 &0.785 & 1.74 &19.0974 &  3508 &1.71 &0.338  &0.459  &0.374  &0.071  \\
          &10 &0.007 &1.30 &0.010 & 7037 &4.194 &1.510 &0.701 & 1.72 &19.0941 &  3430 &1.87 &0.357  &0.486  &0.377  &0.069  \\
          &11 &0.007 &1.32 &0.010 & 7122 &4.195 &1.519 &0.727 & 1.73 &19.0970 &  3491 &1.77 &0.363  &0.495  &0.376  &0.071  \\
\hline
\end{tabular}
\end{table*}
\newpage

\begin{table*}
\centering
\caption{\label{t3}Fundamental parameters.}
\begin{tabular}{lccccc}
\hline\hline
Parameters                      &TIC 65138566            &TIC 139729335 \\
\hline
$M$ ($M_{\odot}$)               &1.36 $\pm$ 0.06         &1.36 $\pm$ 0.06\\
$Z$                             &0.005 $\pm$ 0.002       &0.005 $\pm$ 0.002\\
$T_{\rm eff}$ (K)               &7494 $\pm$ 428          &7560$^{+323}_{-523}$ \\
$\log g$ (cgs)                  &4.17 $\pm $ 0.01        &4.20 $\pm$ 0.01\\
$R$ ($R_{\odot}$)               &1.58 $\pm$ 0.03         &1.54 $\pm$ 0.03\\
$\log (L/L_{\odot})$            &0.85 $\pm$ 0.12        &0.84 $\pm$ 0.14 \\
$\tau_0$ (hr)                   &1.82 $\pm$ 0.03         &1.76 $\pm$ 0.04\\
Age (Gyr)                       &1.73 $\pm$ 0.28         &1.65 $\pm$ 0.25\\
$X_{\rm c}$                     &0.28 $\pm$ 0.05         &0.31 $\pm$ 0.05 \\
$\bar{\rho}/\bar{\rho}_{\odot}$ &0.344 $\pm$ 0.003       &0.373 $\pm$ 0.004 \\
$R_{\rm conv}/R$                &0.068 $\pm$ 0.002       &0.071 $\pm$ 0.003 \\

\hline
\end{tabular}
\end{table*}


\end{document}